\documentclass[twocolumn,tighten]{aastex631}

\usepackage{graphicx}
\usepackage{bm}
\usepackage{url}
\usepackage{threeparttable}
\usepackage{amsmath,amssymb}
\usepackage{xspace}
\usepackage{ulem}
\usepackage{subfigure}

\newcommand{\nustar}{\textit{NuSTAR}\xspace}
\newcommand{\xmm}{\textit{XMM-Newton}\xspace}

\newcommand{\ergs}{erg s$^{-1}$} 
\newcommand{\lopt}{$\lambda L_{\lambda,\rm 5100}$}
 
\shorttitle{Co-evolution and nuclear structure of POX 52} 
\shortauthors{Kawamuro et al.}

\begin{document}
\title{
  Co-evolution and Nuclear Structure in the Dwarf Galaxy POX 52 Studied by Multi-wavelength Data From Radio to X-ray
}

\correspondingauthor{Taiki Kawamuro}
\email{taiki.kawamuro@riken.jp}

\author{Taiki Kawamuro}
\altaffiliation{RIKEN Special Postdoctoral Researcher} 
\affil{RIKEN Cluster for Pioneering Research, 2-1 Hirosawa, Wako, Saitama 351-0198, Japan}

\author{Claudio Ricci}
\affil{Instituto de Estudios Astrof\'isicos, Facultad de Ingenier\'ia y Ciencias, Universidad Diego Portales, Av. Ej\'ercito Libertador 441, Santiago, Chile}
\affil{
Kavli Institute for Astronomy and Astrophysics, Peking University, Beijing 100871, China}

\author{Satoshi Yamada}
\altaffiliation{RIKEN Special Postdoctoral Researcher} 
\affil{RIKEN Cluster for Pioneering Research, 2-1 Hirosawa, Wako, Saitama 351-0198, Japan}

\author{Hirofumi Noda}
\affil{Department of Earth and Space Science, Graduate School of Science, Osaka University, 1-1 Machikaneyama, Toyonaka, Osaka 560-0043, Japan}

\author{Ruancun Li} 
\affil{
Kavli Institute for Astronomy and Astrophysics, Peking University, Beijing 100871, China}
\affil{Department of Astronomy, School of Physics, Peking University, Beijing 100871, China}

\author{Matthew J. Temple}
\affil{Instituto de Estudios Astrof\'isicos, Facultad de Ingenier\'ia y Ciencias, Universidad Diego Portales, Av. Ej\'ercito Libertador 441, Santiago, Chile}

\author{Alessia Tortosa}
\affil{Instituto de Estudios Astrof\'isicos, Facultad de Ingenier\'ia y Ciencias, Universidad Diego Portales, Av. Ej\'ercito Libertador 441, Santiago, Chile}

\begin{abstract}

The nearby dwarf galaxy POX\,52 at $z = 0.021$ hosts an active galactic nucleus (AGN) with a black-hole (BH) mass of $M_{\rm BH} \sim 10^{5-6}\,M_\odot$ and an Eddington ratio of $\sim$ 0.1--1. This object provides the rare opportunity to study both AGN and host-galaxy properties in a low-mass highly accreting system. To do so, we collected its multi-wavelength data from X-ray to radio. First, we construct a spectral energy distribution, and by fitting it with AGN and host-galaxy components, we constrain AGN-disk and dust-torus components. Then, while considering the AGN-disk emission, we decompose optical \textit{HST} images. As a result, it is found that a classical bulge component is probably present, and its mass ($M_{\rm bulge}$) is consistent with an expected value from a local relation. Lastly, we analyze new quasi-simultaneous X-ray (0.2--30\,keV) data obtained by \textit{NuSTAR} and \textit{XMM-Newton}. The X-ray spectrum can be reproduced by multi-color blackbody, warm and hot coronae, and disk and torus reflection components. Based on this, the spin is estimated to be $a_{\rm spin} = 0.998_{-0.814}$, which could suggest that most of the current BH mass was achieved by prolonged mass accretion. Given the presence of the bulge, POX\,52 would have undergone a galaxy merger, while the $M_{\rm BH}$--$M_{\rm bulge}$ relation and the inferred prolonged accretion could suggest that AGN feedback occurred. Regarding the AGN structure, the spectral slope of the hot corona, its relative strength to the bolometric emission, and the torus structure are found to be consistent with Eddington-ratio dependencies found for nearby AGNs.

\end{abstract}

\keywords{galaxies: active -- galaxies: individual (POX 52) -- X-rays: galaxies} 

\section{Introduction}

Supermassive black holes (SMBHs) heavier than a million solar masses are believed to be ubiquitously present at the center of the massive galaxies ($> 10^{9-10}\,M_\odot$, where $M_\odot$ is the solar mass) \citep[e.g.,][]{Gre12,Mil15,Cha23}. Various correlations have been found between galaxies and their SMBHs, such as the one between the bulge and SMBH masses \citep[e.g.,][]{Mag98,Geb00,Mar03,Gul09,Kor13}, and it has been considered that galaxies and SMBHs have co-evolved. Many studies have been conducted to understand the physical mechanisms responsible for this co-evolution \citep[e.g.,][ and references therein]{Fab12,Kin15,Har18}, especially in massive systems. An often hypothesized scenario is that a galaxy merger ignites star formation (SF) and mass accretion onto an SMBH, and, at some point, the resultant active galactic nucleus (AGN) blows gas out of the system in the form of outflows, preventing further star-formation (SF) and SMBH growth \citep[e.g.,][]{Hop08,Hop10,Boo09}. In this scenario, some simulations successfully reproduced the correlations that have actually been observed \citep[e.g.,][]{DiM05,Cro06}. 
In addition to the interplay between the SF and AGN, it has also been proposed that successive mergers are important as they may reduce the scatter around correlations with the SMBH mass \citep{Kor13}. 
However, it is not well understood whether these mechanisms are plausible for less massive systems, or dwarf galaxies with stellar masses of $M_{\rm star} \lesssim 10^{9.5}\,M_\odot$. 

In considering the growth of dwarf galaxies, many studies have  focused on stellar feedback, such as winds, supernovae (SNe), and re-ionization originating in star-forming regions \citep[e.g.,][]{Lar74,Efs92,Fit17}. 
The stellar feedback hypothesis has been successful in explaining discrepancies between what the standard lambda cold dark matter ($\Lambda$CDM) model predicts and the actual properties of dwarf galaxies \citep[e.g., substructure, core-cusp, too-big-to-fail, and satellite-plane problems; e.g.,][]{Moo99,Kro05,DeB10,Boy11}. On the other hand, AGN feedback has often been ignored, due to various possible reasons, such as the possibility that there is no black hole (BH) that can affect surroundings, the depletion of gas to feed the primary BH due to the stellar feedback, and the difficulty for the BH to sink to the central dense-gas region to grow due to the shallow gravitational potential of the galaxy.

While these assumptions have been used for a long time, recently, an increasing number of works have discussed the importance of AGN feedback theoretically and observationally. 
By analytically comparing the outflow properties driven by the AGN and SNs, \cite{Das18} showed that AGN feedback is generally more efficient than the SN feedback \cite[see also][]{Sil17}. Cosmological simulations have also shown that the AGN feedback is capable of affecting host galaxy properties \cite[e.g.,][]{Kou21,Kou22}. 
Several pieces of observational evidence have supported this idea.
An increasing number of AGNs have been found in 
dwarf galaxies 
\cite[e.g.,][and references therein]{Gre04,Gre07,Gre20}. 
\cite{Pen18} found ionized gas that does not follow the dynamical structure of the stellar component and suggested that it is perhaps due to the AGN 
\cite[][]{Man19,Mez20}. Also, \cite{Liu20_dwarf_ifu} detected ionized gas outflows from about 90\% of AGN-hosting dwarf galaxies, and 
suggested that the outflows are likely due to the AGNs. 
Finally, it was found that dwarf galaxies together with massive systems form a $M_{\rm BH}$--$M_{\rm bulge}$ relation, which could suggest the co-evolution even in the less massive systems \cite[e.g.,][]{Sch19,Gre20}.

Studying AGNs in dwarf galaxies is important not only in the above context, but also for revealing the properties of intermediate-mass BH (IMBH) AGNs \citep[i.e., $M_{\rm BH} \sim 10^{3-6}\,M_\odot$; e.g.,][]{Mez17}. 
In fact, it is still unclear whether IMBH AGNs have similar nuclear structures (e.g., the corona, accretion disk, and torus) to those of massive AGNs and whether they follow the same observational trends as suggested for the massive AGNs. Ultimately, 
the understanding of IMBH AGNs is expected to play an important role in linking the accretion physics between stellar-mass BH and SMBHs. Furthermore, in terms of application,
if constrained, its spectral energy distribution (SED) will serve as a useful template for discussing the strategy to detect IMBH AGNs at different redshifts, including the seeds of quasars at high redshifts of $\gtrsim$ 6 \citep{Cru23}. The 
search for the quasar progenitors with various observatories is being actively discussed \cite[e.g.,][]{Val18,Mar20,Gri20}.

In this paper, toward a better understanding of the co-evolution and 
AGN structures in less massive systems, we present our observational study of the dwarf galaxy POX\,52 located at (R.A., Decl.) = (12h02m56.913s, $-$20d56m02.66s) 
in J2000.0 (e.g., \citealt{Kun81}). 
POX 52 is in the nearby Universe of $z = $ 0.021, or $D = 93$ Mpc, adopting a $\Lambda$CDM cosmology with $H_0$ = 70 km s$^{-1}$ Mpc$^{-1}$, $\Omega_{\rm m} = 0.3$, and $\Omega_\Lambda = 0.7$, and is known to host a rapidly growing type-1 IMBH AGN with a BH mass of $M_{\rm BH} \approx 2$--$4\times10^{5}\,M_\odot$ and an Eddington ratio of $\lambda_{\rm Edd} \approx$ 0.2--0.5 \citep[][]{Tho08}.
The mass was derived by the single-epoch method, and 
the bolometric luminosity necessary for the Eddington ratio was 
derived by integrating an SED roughly fitted to observed data from the radio to the X-ray. 
The galaxy, thus, provides us with an invaluable opportunity to reveal both AGN and host-galaxy properties precisely. 
This single-object study is complementary to studies using large samples, given that 
we reveal the nature of the IMBH AGN in quite detail. 

We collected multi-wavelength data, including newly obtained 
broadband X-ray data by \textit{NuSTAR} and \textit{XMM-Newton} quasi-simultaneous observations. 
Our analysis based on these data can be divided into three parts essentially and proceeded as follows. First, we constructed an SED from infrared (IR) to ultra-violet (UV) and decomposed it into AGN and host-galaxy components using the Code Investigating GALaxy Emission (CIGALE) code \citep{Boq19}\footnote{https://cigale.lam.fr/} (Section~\ref{sec:sed}). The AGN components were the radiation from the accretion disk and that from the dust torus. Next, we spatially resolved high-resolution Hubble Space Telescope (\textit{HST}) images into 
AGN and host-galaxy components (Section~\ref{sec:galfit}). Here, in order to avoid the degeneracy of the AGN and host-galaxy components, the AGN component was included so that it was consistent with the SED result. Finally, we analyzed a broadband X-ray spectrum obtained with \textit{NuSTAR} and \textit{XMM-Newton} (Section~\ref{sec:xspec}). 
Our fits to the spectrum took into account the reflection component expected from the dust torus constrained by the SED decomposition, enabling us to reduce degeneracy with other spectral components. In summary, starting from the SED results, we reveal the properties of the host galaxy and the AGN while ensuring no inconsistencies in all the used data.

This paper is organized as follows. 
We present the data used and how these were reduced and/or reprocessed  in Section~\ref{sec:obsdata}. 
Next, as described previously, Sections~\ref{sec:sed}, \ref{sec:galfit}, and \ref{sec:xspec} 
present our SED, imaging, and X-ray data analyses, respectively. We revisit the IMBH mass estimate in Section~\ref{sec:bh}. 
Using our new results, we discuss the evolution of POX 52 and the AGN structure, and summarize our AGN SED model in Section~\ref{sec:dis}. 
Our summary of this paper is finally presented in Section~\ref{sec:sum}. 
Unless otherwise noted, errors are quoted at the 1$\sigma$ confidence level for a single parameter of interest.

\section{Observational Data}\label{sec:obsdata}

\subsection{Data used for constructing an SED}

\subsubsection{\textit{XMM-Newton} Optical and UV Data}

Using \textit{XMM-Newton} \citep{Jan01}, we obtained X-ray, optical, and UV data (ID = 0890430101) with an exposure of $\approx$ 20 ks between 2021 December 30 and 31. The data were analyzed with the Science Analysis Software (SAS) version 20.0.0 by following the \textit{XMM-Newton} ABC guide\footnote{https://www.cosmos.esa.int/web/xmm-newton/sas-threads}. 
The latest calibration files were used.
The Optical Monitor (OM) onboard \textit{XMM-Newton} obtained data in the Image+Fast mode from optical to UV using all the available filters ($V$, $B$, $U$, UVW1, UVM2, and UVW2). 
The obtained data in the Image and Fast modes were reprocessed 
with \texttt{omichain} and \texttt{omfchain}, respectively. 
By inspecting the data in the Image mode, it was found that the background levels around POX 52 were 
enhanced in the images in the $V$, $B$, $U$, and UVW1 filters because of stray light caused by the reflection of a star outside the field of view and bad pixels in the source regions. 
Thus, we decided not to use the photometric data in these four bands. For the two remaining bands (UVM2 and UVW2), we measured source count rates with the SAS task \texttt{omphotom}; we fitted a PSF to the image while considering the background level inferred from an annulus between five and ten pixels. 
The count rates were then converted into flux densities based on the factors suggested by the \textit{XMM-Newton} User Guide\footnote{https://xmm-tools.cosmos.esa.int/external/xmm\_user\_support\\/documentation/sas\_usg/USG/ommag.html}. 
The obtained flux densities were then corrected for dust extinction by adopting the extinction law of \cite{Car89} with $R_{\rm V}$ = 3.1 and $E(B$--$V)$ = 0.05\footnote{The https://irsa.ipac.caltech.edu/applications/DUST/}. 
The resultant flux densities in the UVM2 and UVW2 bands are 
$0.21\pm0.01$ mJy and $0.25\pm0.02$ mJy, respectively. 
We mention that this extinction correction was applied to the other photometric data from UV to near-IR wavelengths as well. 
Regarding the Fast-mode data, a lightcurve was created in each filter and was examined to assess whether the emission varied 
significantly during each $\approx$ 3 ks exposure by fitting a constant model using the chi-square method. We found that no significant variations were present, given the $p$-values $>0.1$ recovered.

\subsubsection{\textit{GALEX} Data} 

We used the \textit{GALEX} GR6/7 Data Release:\dataset[10.17909/T9H59D]{\doi{10.17909/T9H59D}}
to cover the FUV (1516\,\AA) and NUV (2267\,\AA) bands \citep{Mar05,Bia17}, 
which was used also by \cite{Tho08}.
The magnitudes derived by adopting an aperture with a radius of 4\arcsec\ were adopted. 
The respective magnitudes are
19.10$\pm$0.14\,mag and 
18.79$\pm$0.07\,mag 
in the AB system. Hereafter, all magnitudes are presented in the same system. 
By considering the aperture correction ($\Delta m = -0.10$ for the FUV band, 
and $\Delta m = -0.12$ for the NUV band) and dust extinction, the resultant flux densities were calculated to be 0.13$\pm$0.02 mJy and 0.20$\pm$0.02\,mJy in the FUV and NUV bands, respectively.

\subsubsection{PanSTARRS Data} 

We utilized PanSTARRS data to add optical flux densities to our SED \citep{Cha16,Fle20}. 
We adopted the Mean-object catalog, released in the PS1 DR2:\dataset[10.17909/s0zg-jx37]{\doi{10.17909/s0zg-jx37}}. 
In the catalog, POX 52 was identified as an extended source. Thus, to estimate the flux of the entire galaxy, we adopted magnitudes derived in the Kron photometry and corrected them by considering that the Kron magnitude 
within 2.5 $\times$ 1st radial moment generally underestimates the entire flux by $\sim$ 10\%\footnote{https://ned.ipac.caltech.edu/level5/March05/Graham\\/Graham2\_6.html}. 
The $g$-, $r$-, $i$-, $z$-, and $y$-Kron magnitudes listed in the catalog are 17.29\,mag, 16.89\,mag, 16.76\,mag, 16.65\,mag, and 16.52\,mag, and the flux densities corrected for the aperture and dust extinction were estimated to be 0.578$\pm$0.002\,mJy, 0.800$\pm$0.004\,mJy, 0.876$\pm$0.008 mJy, 0.948$\pm$0.010\,mJy, and 1.06$\pm$0.01\,mJy, respectively.

\subsubsection{2MASS Data} 

To cover the near-IR band, we used the photometry data of the Two Micron All Sky Survey (2MASS) All-Sky Point Source Catalog \citep{Skr06} in the same manner as \cite{Tho08}. 
We note that POX 52 was not listed in the Extended Source Catalog. 
The observed magnitudes in the $J$, $H$, and $Ks$ bands (i.e., 1.24\,$\mu$m, 1.66\,$\mu$m, and 2.16\,$\mu$m) are 15.72$\pm$0.08\,mag, 14.96$\pm$0.08\,mag, and 14.46$\pm$0.08\,mag. 
These magnitudes were derived by a profile-fitting method, and the normalizations were adjusted to include the entire fluxes\footnote{https://irsa.ipac.caltech.edu/data/2MASS/docs/releases/\\allsky/doc/sec4\_4c.html}. 
By correcting the observed values for dust extinction, the flux densities in the $J$, $H$, and $Ks$ bands were estimated to be 0.86$\pm$0.06 mJy, 1.09$\pm$0.08 mJy, and 1.12$\pm$0.09 mJy, respectively.

\subsubsection{\textit{WISE} Data} 

In the IR band, we used \textit{WISE} photometric data in the four bands: 3.4\,$\mu$m, 4.6\,$\mu$m, 12\,$\mu$m, and 22\,$\mu$m ($W1$, $W2$, $W3$, and $W4$, respectively). POX 52 was detected in all bands in the ALLWISE Source Catalog \citep{Wri10,Mai11},
and its \textit{WISE} data were used in a low-mass AGN study by \cite{Mar17}. 
As no bad flags were raised, we used the listed photometric data: 13.82$\pm$0.03\,mag, 13.23$\pm$0.03\,mag, 10.07$\pm$0.06\,mag, and 7.49$\pm$0.13\,mag in the $W1$, $W2$, $W3$, and $W4$ bands, respectively.
These were derived with a profile-fitting method, appropriate for a point source and POX 52 as well, because POX 52 is identified as a point source in the catalog (\texttt{ext\_flg} = 0). 
We then converted them into flux densities by assuming coefficients\footnote{Different coefficients do not affect the conversions much (i.e., $\lesssim$ 10\%, see \\ https://wise2.ipac.caltech.edu/docs/release/allsky/expsup\\/sec4\_4h.html} appropriate for a power-law function of $S_\nu \propto \nu^{-2}$ ($S_\nu$ represents flux density). The flux densities were found to be $0.92\pm0.02$ mJy, $0.87\pm0.02$ mJy, $2.90\pm0.16$ mJy, and $8.5\pm1.0$ mJy, in the $W1$, $W2$, $W3$, and $W4$ bands, respectively. 

\subsubsection{\textit{Spitzer} Data} 

To put stronger constraints on the emission in the IR band, we also used \textit{Spitzer}/IRS spectral data provided by the CASSIS catalog \citep{Leb11,Leb15}. 
The spectrum was investigated by \cite{Hoo17}, 
and many emission lines were identified by them. Among the lines, [Ne V] at $\lambda =$ 14.32 $\mu$m and 24.32 $\mu$m and [O IV] at $\lambda = $ 25.89 $\mu$m would be associated with AGN radiation given photons with energies $\gtrsim 50$ eV are necessary to produce them.  
In our SED analysis, we used the CIGALE code \citep[][Section~\ref{sec:sed}]{Boq19}, and this code does not take into account AGN-related emission lines. Thus, we excluded the wavelength bands around the 
[Ne V] and [O IV] lines. Specifically, the 14.4--14.9 $\mu$m and 23.6--27.8 $\mu$m ranges were excluded. 
Also, to reduce the computational cost, we rebinned the spectrum so that each bin size was 0.2 $\mu$m. 

We mention here two points: the high-ionization potential lines seen in the IRS spectrum are unlikely to contribute much to the \textit{WISE} W3 and W4 fluxes, and the \textit{Spitzer}/IRS spectrum would accommodate the entire galaxy. 
For the first point, we calculated the equivalent widths (EWs) of [Ne V] 14.32 $\mu$m, [Ne V] 24.32 $\mu$m, and [O IV] 25.89 $\mu$m lines to be 
$\sim$ 0.2 $\mu$m,
$\sim$ 0.2 $\mu$m, and 
$\sim$ 1 $\mu$m. Here, we used the line fluxes measured by \cite{Spo22}. 
It is thus clear that the [Ne V] 14.32 $\mu$m line contributes little to the flux in 
the W3 band, whose effective bandwidth is $\approx$ 5.5 $\mu$m. On the other hand, 
in the W4 band with the effective bandwidth of $\approx$ 4.1 $\mu$m, the [O IV] 25.89 $\mu$m line, in particular, is relatively large (i.e., $\sim$ 30\%). However, given that the response is significantly low around the [O IV] line, its contribution to the flux can be safely ignored. 
For the second point, Figure~\ref{fig:sed} 
shows the W3 and W4 flux densities as well as the IRS spectrum. It is clearly seen that the 
two \textit{WISE} fluxes and the adjacent IRS fluxes are comparable. As the \textit{WISE} fluxes were measured for the entire galaxy, we concluded that no aperture correction was needed for the IRS spectrum. 

\subsubsection{VLA Data} 

Very Large Array (VLA) data in the $S$ and $C$ bands are available.
The $S$-band ($\approx$ 3\,GHz) data were taken in the course of the VLA sky survey 
\citep[VLASS;][]{Gor20,Gor21}. 
The survey data were already analyzed by a VLASS team, and reconstructed images are available in the VLASS website\footnote{https://www.cadc-ccda.hia-iha.nrc-cnrc.gc.ca/en/vlass/}. By examining an image encompassing POX 52, we found no significant emission from the source position. Thus, 
using noise in an off-source region, the upper limit at a $3\sigma$ level was estimated to be 0.6 mJy beam$^{-1}$, where beam size is $\sim$ 2\farcs5. The corresponding upper-limit luminosity is $\lesssim 2\times10^{37}$ erg s$^{-1}$. 

The $C$-band ($\approx$ 5\,GHz) data were obtained by a pointing observation, and were analyzed by \cite{Tho08}, who found that no significant emission was detected by the $C$-band observation. The 3$\sigma$ upper limit was estimated to be 0.08\,mJy, or $4\times10^{36}$\,erg\,s$^{-1}$, by the authors, and we used this limit.

\subsection{\textit{HST} Data}

We used \textit{HST} data with the goal of revealing stellar components of the host galaxy by fitting the images. POX 52 was observed with the Advanced Camera for Surveys (ACS) and the High Resolution Channel (HRC) on 2004 November 19 and 18. 
Two optical filters of F435W and F814W were adopted in the respective days, and the data were analyzed by \cite{Tho08} in the past.
We used the data stored in the archive of Mikulski Archive for Space Telescopes (the data are available at MAST:\dataset[10.17909/J2EW-NH46]{\doi{10.17909/J2EW-NH46}})\footnote{https://mast.stsci.edu/search/ui/\#/hst}. 
The entire exposure in each filter was split into five intervals, each having calibrated and flat-fielded images. 
We combined the five calibrated images into one with an \textit{HST} pipeline of \texttt{astrodrizzle}, where \texttt{final\_scale} was set to 0\farcs025. We again run \texttt{astrodrizzle} to produce a noise image relevant to each image by setting \texttt{final\_wht\_type} to \texttt{ERR} \citep[e.g.,][]{Flo08,Kok20}. The noise images were necessary to perform the GALFIT fitting \citep[][]{Pen02,Pen10} in Section~\ref{sec:galfit}. 
The total exposures of the combined data in the F435W and F814W filters are $\approx$ 2.6\,ks and $\approx$ 2.5\,ks, respectively. 
In addition to producing the observed images and the relevant noise images, we generated a PSF image in each filter using a PSF modeling software of TinyTim \citep{Kri11}. These images were also necessary for the image fit.

\begin{figure*}
  \centering
  \includegraphics[scale=0.55]{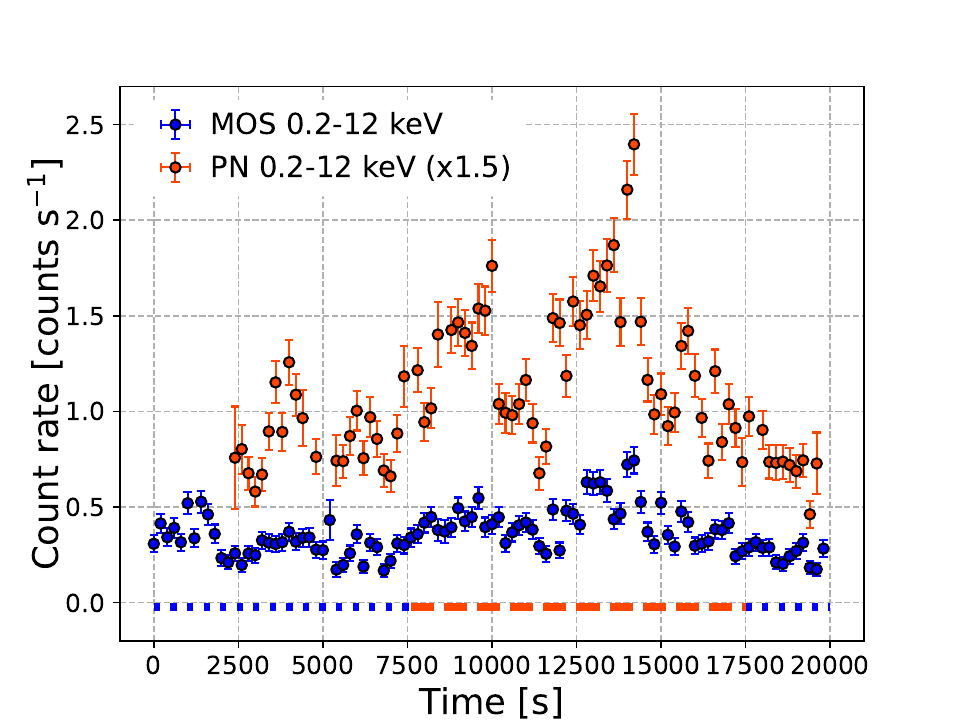}
  \includegraphics[scale=0.55]{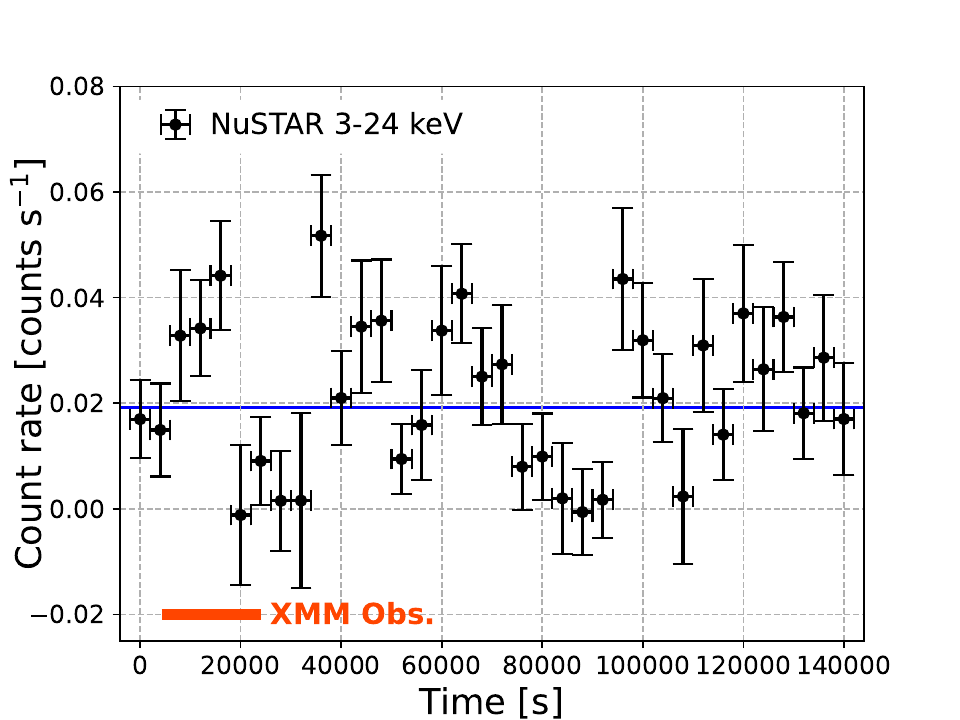} 
  \caption{
  Left: 
  \textit{XMM-Newton} MOS (blue) and PN (orange) lightcurves in the 0.2--12 keV band. 
  The adopted binsize is 200 sec. 
  For clarity, the PN lightcurve is increased by a factor of 1.5. 
  The blue dotted and orange dashed lines indicate fainter and brighter phases, we defined in Section~\ref{sec:xspec_var}. 
  Right: \textit{NuSTAR} lightcurve in the 3--24 keV band, created with a binsize of 4 ks.
  The blue solid line corresponds to the average value, and the orange solid line indicates when the \textit{XMM-Newton} observation was executed.   
  }\label{fig:xlc} 
\end{figure*}

\begin{figure}
  \centering
  \includegraphics[scale=0.5]{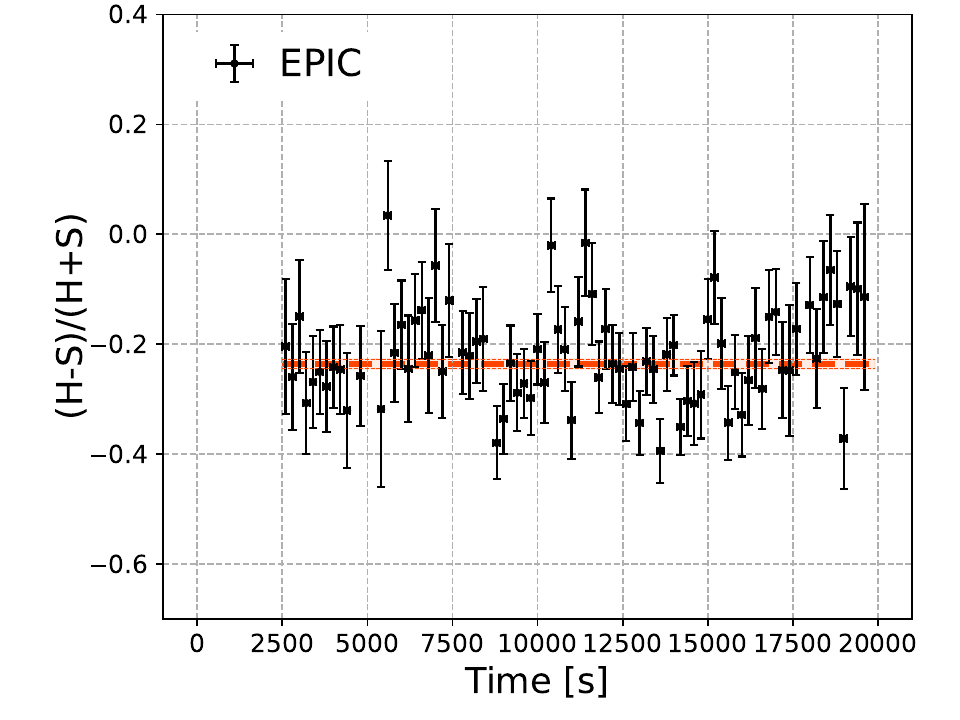}  \vspace{-0.3cm}
  \caption{
  \textit{XMM-Newton}/EPIC hardness ratios, defined as (H-S)/(H+S), where S and H are count rates in the 0.2--1 keV and 1--12 keV bands. 
  The adopted binsize is 200 sec. 
  The orange line indicates the constant model fitted to the data. 
  }\label{fig:xhr} 
\end{figure}

\subsection{X-ray Spectra and Light-curves} 

\subsubsection{\textit{XMM-Newton} X-ray Data}\label{sec:xmm}

The raw \textit{XMM-Newton} MOS and PN data in the X-ray band were initially reprocessed using \texttt{emproc} and \texttt{epproc}, respectively. 
To filter periods with high background counts, we created lightcurves of the MOS1 and MOS2 above 10\,keV 
with PATTERN=0 (single events) and a PN one in the 10--12\,keV band with the same pattern selection. For the MOS1 and MOS2 data, we adopted the threshold of 0.30\,counts\,s$^{-1}$ to define their good time intervals (GTIs). 
For the PN data, by inspecting the PN lightcurve, we left only the period when the count rates were less than 0.45\,counts\,s$^{-1}$ as the GTI. Because the PN data were affected by a strong background flare and the tail of the flare could not be removed completely with the simple filter described above, we also excluded the events around the flare. Finally, by applying a PATTERN selection of $\leq$ 4 plus FLAG=0 and $\leq$ 12 to the PN and MOS data, respectively, we obtained cleaned data. 
Source events were then extracted from circular regions of radii 30\arcsec and 40\arcsec, respectively. Similarly, background count rates were estimated from off-source regions with radii of 30\arcsec and 120\arcsec, respectively. 
The background-subtracted MOS and PN lightcurves are shown in Figure~\ref{fig:xlc}.
Based on the count rate in each detector, we confirmed that pile-up effects are negligible\footnote{https://xmm-tools.cosmos.esa.int/external/xmm\_user\_support\\/documentation/uhb/epicmode.html\#3775}. 
The hardness ratios shown in Figure~\ref{fig:xhr} were calculated in each bin by combining the MOS and PN data (i.e., periods when good PN data were unavailable were not considered). No significant variability of the hardness ratios was confirmed by fitting a constant model and obtaining a $p$-value of $> 0.01$. 
Spectra were extracted from the same source and background regions. The MOS1 and MOS2 spectra were combined into one using \texttt{epicspeccombine}. We then binned the source spectra to have a minimum of one count per bin. Response files were produced using the SAS tasks \texttt{arfgen} and \texttt{rmfgen}. Particularly in generating the arf files, a flag was raised for \texttt{applyabsfluxcorr} to correct the effective area to remove residuals between \textit{NuSTAR} and \textit{XMM-Newton} spectra.

\subsection{\textit{NuSTAR} Data}\label{sec:nustar}

\textit{NuSTAR} \citep{Har13} observed the target for $\approx$ 80 ks quasi-simultaneously with \textit{XMM-Newton} (ID = 60701047002). The \textit{NuSTAR} observation started around 5 ks later than the \textit{XMM-Newton} one. 
Following the ``\textit{NuSTAR} Analysis Quickstart Guide", we used the standard \texttt{nupipeline} script in the NuSTARDAS version v2.1.2 for the reprocessing. We set SAAMODE=optimized, since the background event rates were slightly elevated around the South Atlantic Anomaly. We used the reduced data for further analyses. We note that although we created different reduced data with more strict options of SAAMODE=strict and TENTACLE=yes, we found that the resultant spectra and lightcurves were similar to those obtained for SAAMODE=optimized. 

We defined the source region as a 40\arcsec-radius circle centered at the optical position, by taking account of the FWHM of the \textit{NuSTAR} PSF ($\approx$18\arcsec). For the background region, we adopted an off-source circular region with a 120\arcsec-radius on the same detector. Then, we produced lightcurves, spectra, and response files using the \texttt{nuproducts} task. The produced lightcurves in the 3--24 keV range are shown in the right panel of Figure~\ref{fig:xlc}. 
Significant variability was found by fitting a constant model to the data and obtaining a $p$-value of 3$\times10^{-4}$. 
It was, however, confirmed that the 
flux level during the \textit{XMM-Newton} 
observation was consistent with that for the entire \textit{NuSTAR} observation within $\sim$ 1$\sigma$. Thus,
Thus, our \nustar spectra, to be fitted with the \xmm spectrum, were 
produced by combining all data in each detector (i.e., FPMA and FPMB). 
Each of the two spectra was then binned using the same approach we adopted for the \textit{XMM-Newton} spectra.

\begin{deluxetable}{cccc}
\tabletypesize{\footnotesize}
\tablecaption{Allowed Parameter Grids in the SED Fitting and Best-fit Parameters \label{tab:cigale}}
\tablewidth{0pt} 
\startdata \vspace{-0.1cm} \\ 
Parameter & Grids & Best \\
\hline \hline 
\multicolumn{3}{c}{delayed SFH}\\
\hline
$\tau_{\rm main}$ [Myr]  & 1,000, 3,000, 6,000 & 3000 \\
age [Myr]                & 2500, 5000, 7500, 10000 & 7500 \\
\hline \hline
\multicolumn{3}{c}{SSP \citep{Bru03}} \\ 
\hline
IMF                      & \cite{Cha03} & ... \\
Metallicity              & 0.004, 0.008, 0.02 & 0.004 \\
\hline \hline
\multicolumn{3}{c}{Nebular emission \citep{Ino11}}\\
\hline
$\log\, U$               & $-2.0$ & ... \\
Metallicity              & 0.004 (fixed to the SSP one) & ... \\
\hline \hline
\multicolumn{3}{c}{Dust attenuation \citep{Cha00}}\\
\hline
$A_{\rm V,ISM}$     & 0.2, 0.4, 0.6, 0.8 & 0.6 \\
\hline \hline
\multicolumn{3}{c}{AGN emission \citep{Sta16}} \\
\hline
$\tau_{\rm 9.7}$             & 3, 5, 9          & 3 \\
$p_{\rm torus}$                          & 1.0              & ... \\
$q_{\rm torus}$                          & 1.0              & ... \\
$\sigma_{\rm torus}$ [$\degr$]  & 20, 10           & 10 \\
$R_{\rm torus}$    & 20               & ... \\
$\theta_{\rm inc}$ [$\degr$] & 20, 30, 40       & 20 \\
$f_{\rm AGN}$                & 0.005, 0.01, 0.5 & 0.01 \\
disk type                    & \cite{Sch05}     & ... \\ \hline \hline
\multicolumn{3}{c}{Dust emission \citep{Dra14}}\\
\hline
$q_{\rm pah}$ & 0.47, 1.12, 2.50 & 1.12 \\ 
$u_{\rm min}$ & 5, 25, 50  & 25 \\ 
$\alpha$      & 2 & ... \\ \hline \hline
\multicolumn{3}{c}{Other Best-fit Parameters}\\ \hline 
SFR          & ... & 0.15$\pm0.01\,M_\odot$ yr$^{-1}$ \\
$M^{\rm old}_{\rm star}$ & ... & $8.3\pm0.4\times10^{8}\,M_\odot$ \\
$M^{\rm young}_{\rm star}$ & ... & $1.4\pm0.1\times10^{6}\,M_\odot$ \\
\enddata
\tablecomments{
For a detailed explanation, we guide readers to the CIGALE papers of \cite{Boq19} and \cite{Yan20}. The second and third columns indicate the parameter grids used to fit the SED and the obtained best-fit parameters, respectively. 
}
\end{deluxetable}

\begin{figure*}
  \centering
  \includegraphics[scale=0.8]{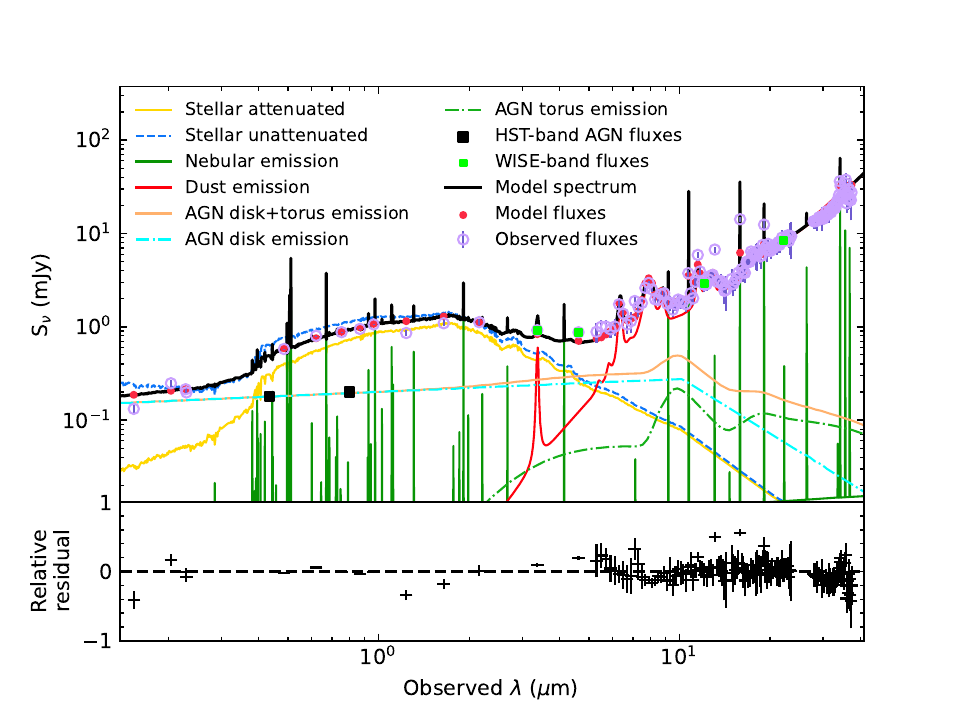}
  \caption{
  SED of POX\,52 (purple open circles), fitted by a model (black line), which includes stellar emission, nebular emission, dust emission originating in the SF, and AGN disk and torus emission. Corresponding lines are denoted in the figure. 
  The black-filled squares indicate the fluxes adopted to model the AGN emission in resolving the \textit{HST} images into AGN and host-galaxy components (see Section~\ref{sec:galfit}). Also, the \textit{WISE}-band fluxes are shown with green squares  so that one can distinguish between the \textit{WISE} and \textit{Spitzer}/IRS data. 
  The lower panel indicates residuals, defined as 
  the difference between the observed and model fluxes divided by the observed one. 
  }\label{fig:sed} 
\end{figure*}

\section{SED analysis}\label{sec:sed}

With the CIGALE code \citep{Boq19}, we decomposed our constructed  SED (Figure~\ref{fig:sed}) consisting of the collected flux densities from the IR to the UV. We did not consider the radio and X-ray data in our SED analysis.
The radio data are later used to discuss whether the upper limits are consistent with the results of our SED analysis, while the X-rays were not considered since they require complex models to take into account all their futures. A detailed analysis of the broad-band X-ray spectrum is reported in Section~\ref{sec:xspec}.  

CIGALE can compute SED models from IR to UV in a self-consistent framework, keeping the energy between the UV/optical and IR wavelengths balanced. 
Among emitting mechanisms and objects whose modes are implemented in CIGALE, we considered SF, stars, AGN disk emission, and AGN dusty torus emission, as listed in Table~\ref{tab:cigale}. 
The SF history was assumed to decrease exponentially with an e-folding time 
\citep[i.e., \texttt{sfhdelayed}; e.g.,][]{Cie15,Cie16}.
To reproduce emission due to SF, we selected the single-stellar-population model of \texttt{bc03} \citep{Bru03}, and the IMF of \cite{Cha03}. 
The standard nebular emission model (\texttt{nebular}) associated with the stellar model was also incorporated. 
To model dust attenuation of the host-galaxy emission, we adopted the model constructed by following the procedure proposed in \cite{Cha00} (\texttt{dustatt\_modified\_CF00}). 
Radiation from old stars (age $> 10$ Myr) is attenuated by dust in the interstellar medium (ISM), while that from younger stars is additionally subject to dust attenuation in a birth cloud (BC). The two dust-attenuation models in the ISM and the BCs are represented by power-laws normalized to the amount of attenuation in the $V$-band, which was left as the free parameter. The power-law indices for the ISM and BC sites were set to the canonical values of $-0.7$ and $-1.3$. 
The amounts of $V$-band attenuation for the young and old stars were linked as $A^{\rm ISM}_{\rm V}/(A^{\rm ISM}_{\rm V} + A^{\rm BC}_{\rm V}) = 0.44$. 
The dust emission from the host galaxy was modeled with  \texttt{dl2014} created by following \cite{Dra14}, which is the refined model of \cite{Dra07}. 
The parameters were set while considering recent works \citep{Bua18,Bua21}. The dust fraction in photo-dissociation regions was fixed to 0.02 \cite[e.g.,][]{Dra07}.
Next, we modeled the AGN torus and disk emission using SKIRTOR 
\citep[i.e., \texttt{skirtor2016};][]{Sta16}, assuming  
a two-phase medium torus model where high-density clumps distribute in a low-density region. 
The parameters are the torus optical depth at 9.7 $\mu$m ($\tau_{9.7}$), the torus density radial parameter ($p_{\rm torus}$), the torus density angular parameter ($q_{\rm torus}$), the angle between the equatorial plane and the edge of the torus ($\sigma_{\rm torus}$), the ratio of the maximum to minimum radii of the torus ($R_{\rm torus}$), the inclination angle ($\theta_{\rm inc}$), and the AGN fraction of the total IR luminosity ($f_{\rm AGN}$). 
Using the parameter settings and grids listed in Table~\ref{tab:cigale}, we fitted the above-described models. The best-fit parameters are also listed in the table.  
For a sanity check, we performed the CIGLAE mock analysis; this analysis initially simulates photometric data based on the best-fit components and the observed photometric uncertainties, and then fits the adopted components to assess whether the best-fit parameters can be recovered even for the simulated data \citep[e.g.,][]{Yan20}. We then confirmed that the best-fit parameters can be surely recovered.

We confirm that the best-fit model is consistent with the properties in the radio and X-ray bands. 
The SFR of POX\,52 is estimated to be $\approx$ 0.15 $M_\odot$ yr$^{-1}$ at $z \approx 0.02$, and 
corresponds to $4\times10^{36}$ erg s$^{-1}$ at 1.4\,GHz via a relation in \cite{Ken12}. 
For an index of 0.8 typical for star-forming galaxies \citep{Tab17}, the corresponding luminosities at 3\,GHz and 5\,GHz were $\approx$ 4--5$\times10^{36}$ \ergs. 
These are consistent with the upper limits constrained in the $S$- and $C$-band VLA observations. 
Regarding the AGN, its radio luminosity should be below $4\times10^{36}$\,erg\,s$^{-1}$ as well. As the 2--10\,keV luminosity of the AGN is $\sim 10^{42}$ erg s$^{-1}$ (Section~\ref{sec:xspec}), the X-ray radio loudness defined as $L_{\rm 1.4}/L_{\rm 2-10}$ should be $< 4\times10^{-6}$. 
This is consistent with observed radio loudnesses of radio-quiet AGNs \citep[][]{Pan15}. In summary, our SED fit does not contradict the radio upper limits and, also, the limits suggest that the AGN is radio quiet. 

The parameters constrained by the SED analysis, particularly, the AGN parameters, played very important roles in decomposing the \textit{HST} images and also in the X-ray spectral analysis. Details are described in Sections~\ref{sec:galfit} and \ref{sec:xspec}.

%

\section{\textit{HST} Image Decomposition}\label{sec:galfit}

\begin{figure*}
  \centering  
  \includegraphics[scale=0.5]{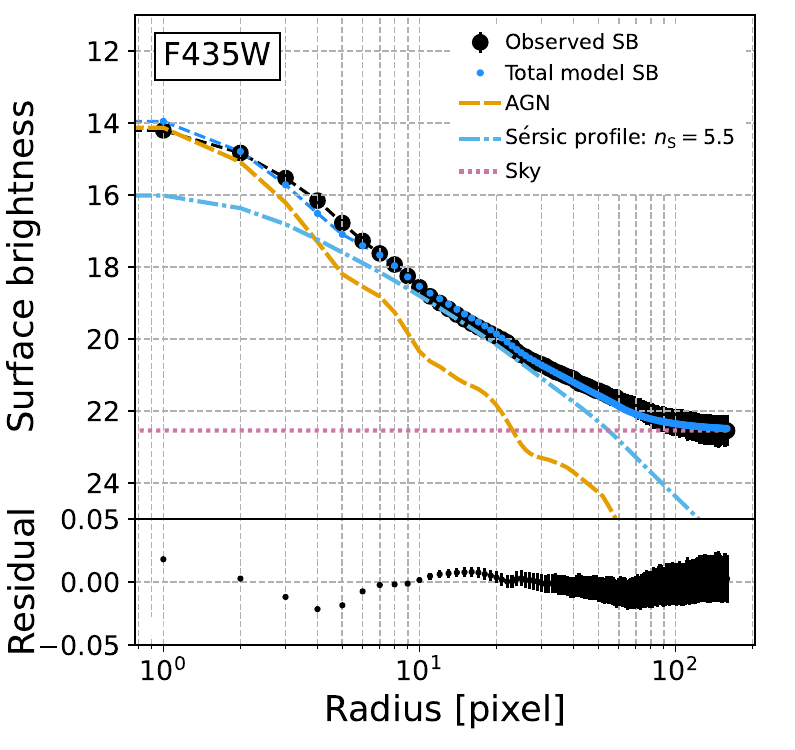}
  \includegraphics[scale=0.5]{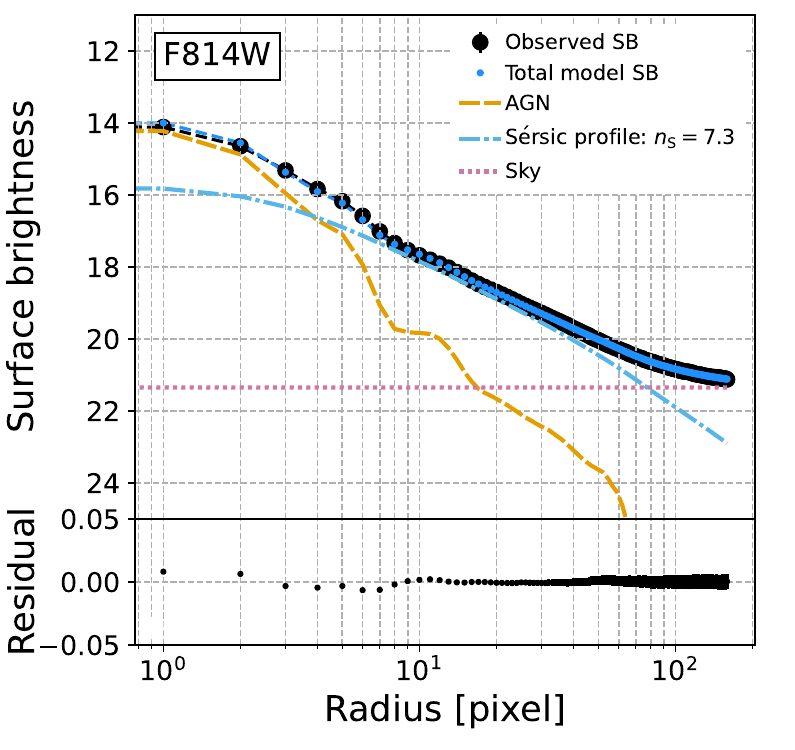}   
  \includegraphics[scale=0.5]{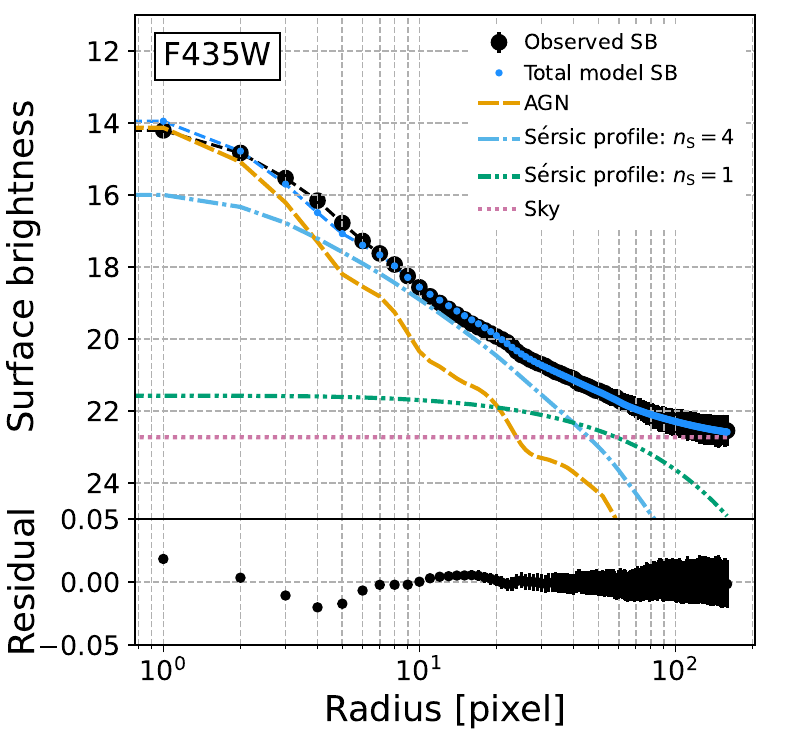}
  \includegraphics[scale=0.5]{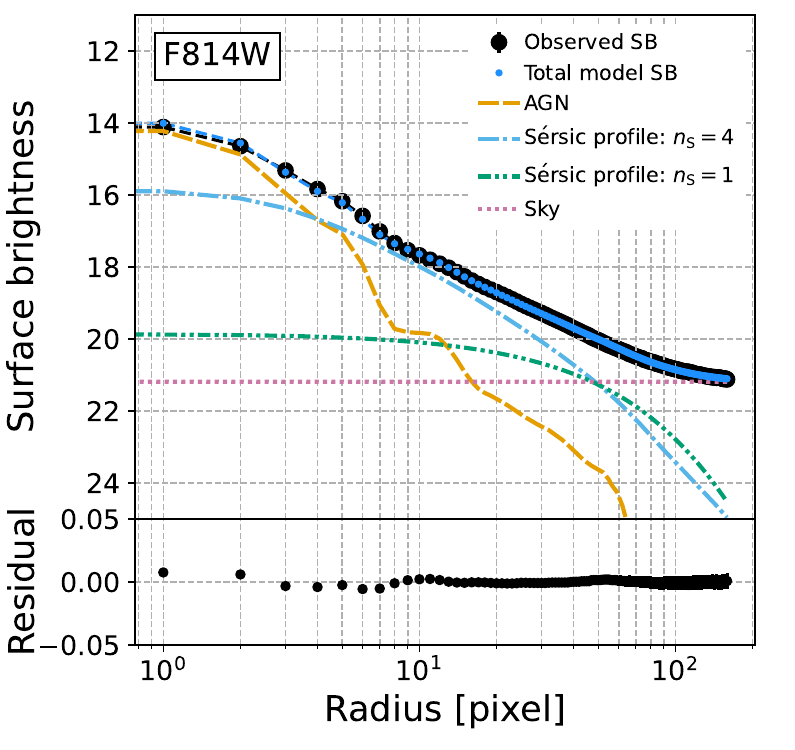} 
  \includegraphics[scale=0.5]{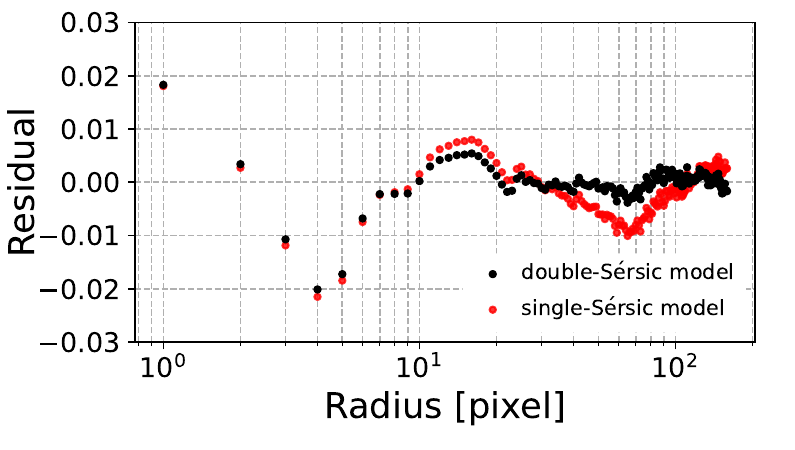} 
  \includegraphics[scale=0.5]{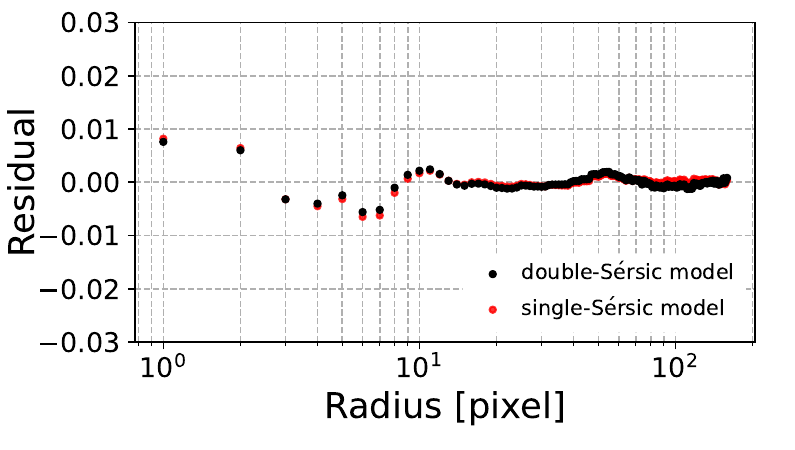}
  \caption{
  Top panels: Surface brightness radial profiles obtained from the \textit{HST} observation with the F435W and F814W filters. 
  Each profile is reproduced by a model consisting of the AGN (i.e., PSF), S\'ersic-profile, and sky components. These were fitted by using the GALFIT code. 
  One pixel corresponds to 0\farcs025 ($\approx$ 11\,pc). 
  Each lower window shows residuals, or (model -- data)/data values. 
  Middle panels: Same as the top panels, but for the double-S\'ersic model, where a S\'ersic profile is added to the model used in the upper panels. 
  Bottom panels: Comparisons of the residuals obtained by the single- and double-S\'ersic profiles in the two filters. For clarity, uncertainties are not shown. 
  }\label{fig:galfit} 
\end{figure*}

\begin{deluxetable}{ccccc}
\tablecaption{Best-fit Parameters of the Fits to the \textit{HST} images\label{tab:galfit}}
\tablewidth{0pt} 
\startdata \vspace{-0.1cm} \\ 
   & Parameter & F435W  & F814W & Units \\ \hline \hline 
  \multicolumn{4}{c}{Point Source (AGN)}\\ \hline
  (1) & $m$ & $18.44^\dagger$  & $18.22^\dagger$  & mag \\ \\ \hline \hline 
  \multicolumn{4}{c}{Single S\'ersic Profile}\\ \hline     
  (2) & $m$                & 18.17           & 16.19 & mag \\ 
  (3) & $n_{\rm S}$        & 5.5             & 7.3    & ... \\ 
  (4) & $R_{\rm e}$        & 16.3            & 120 & pix \\   
  (5) & $b/a$              & 0.88            &  0.88  & ... \\   
  (6) & $PA$               & 1.5             &  37.4 & \arcdeg\\  \\ \hline \hline
  \multicolumn{4}{c}{First S\'ersic Profile}\\ \hline
  (7) & $m$                 & 18.53           & 17.37 & mag \\ 
  (8) & $n_{\rm S}$                 & 4$^{\dagger}$  & 4$^{\dagger}$ & ... \\ 
  (9) & $R_{\rm e}$         & 8.8            & 19.7 & pix \\   
  (10) & $b/a$               & 0.84            & 0.97  & ... \\     
  (11) & $PA$                & $-$2.3            & 3.5   & \arcdeg \\   \hline     
  \multicolumn{4}{c}{Second S\'ersic Profile}\\ \hline   
  (12) & $m$                 & 18.96           & 17.97 & mag \\ 
  (13) & $n_{\rm S}$                 & 1$^{\dagger}$  & 1$^{\dagger}$ & ... \\ 
  (14) & $R_{\rm e}$         & 93.4           & 72.1 & pix \\   
  (15) & $b/a$              & 0.79            &  0.68  & ... \\   
  (16) & $PA$               & 55.3            &  41.7 & \arcdeg\\  \hline 
  (17) & $B/T$ & 0.60 & 0.63 & ... \\ 
\enddata
\tablecomments{
(1) Total magnitude of the point-source model. 
(2,3,4,5,6) Total magnitude, S\'ersic index, half-light radius (1 pix = 0\arcsec025 $\approx$ 11 pc), axis ratio, and position angle of a S\'ersic model fitted to the data together with the point-source and sky background components (i.e., single-S\'ersic model). 
(7)--(16) S\'ersic-function parameters for the double-S\'ersic model. 
(17) Bulge-to-total flux ratio for the double-S\'ersic model. 
$^{\dagger}$These parameters were fixed in the GALFIT fits. 
}
\end{deluxetable}

We decomposed each of the \textit{HST} images taken with the two filters (F435W and F814W) into AGN and host-galaxy components using the image fitting code of GALFIT \citep{Pen02,Pen10}. 
First, as a simple model, we considered a point-source component and a S\'ersic function to represent AGN and stellar emission, respectively. 
Adding a sky background component furthermore, we fitted the \textit{HST} images while leaving all possible parameters as free parameters. The result was that the total magnitudes of the point source were largely different between the two adjacent bands (i.e., $\Delta m \approx 1.5$). 
To avoid this probably unreasonable result, in each band, we fixed the AGN contribution by referring to the AGN disk flux density constrained by the SED analysis. 
The actual magnitudes we adopted were 18.44\,mag and 18.22\,mag in the F435 and F814W filters, respectively. Also, the AGN position was fixed to the peak in each image. Under these conditions, the \textit{HST} images in the F435W and F814 filters seemed to be reproduced well with the S\'ersic indices ($n_{\rm S}$) of 5.5 and 7.3, respectively, as shown in the top panels of Figure~\ref{fig:galfit}. The best-fit parameters of the fitted single-S\'ersic models are listed in Table~\ref{tab:galfit}. 
Although the indices are slightly higher than the past estimates of $\sim 4$ based on ground-based images by \cite{Bar04} and those by \cite{Tho08} using the same \textit{HST} images, all estimated values ($n_{\rm S} \gtrsim 4$) could suggest that the stellar emission is dominated by a bulge \citep[e.g.,][]{Mee15}.

We furthermore added a S\'ersic function to the above model to assess whether there was an additional component, such as a disk component, occasionally presumed to have $n_{\rm S} \sim 1$ \citep[e.g.,][]{All06,Men14}. 
If none or only one of the S\'ersic indices is fixed, 
an unreasonable result is obtained; for example, in the case of fixing only the index of one S\'ersic component to $n_{\rm S} = 1$, the other S\'ersic index was constrained to be $n_{\rm S} \sim$ 2 for the F435W data, and a much larger value of $n_{\rm S} \sim$ 6 was obtained for the F814W data. 
Therefore, we fixed one of the two S\'ersic indices to 1 and the other to 4. 
The results of fitting this double-S\'ersic model to 
the surface brightness profiles are shown in the middle panels of Figure~\ref{fig:galfit}, and the best-fit parameters are listed in Table~\ref{tab:galfit}. 
To compare the goodness-of-fits of the double-S\'ersic and single-S\'ersic models, in the lower panels of Figure~\ref{fig:galfit}, 
we plot the residuals, or (model -- data)/data, left by the two models for each filter. 
As seen for the F435W data more clearly, the double-S\'ersic model seems to fit the data better than the single-S\'ersic one. This is suggested by a decreased chi-square value by an order of 1000 for the addition of six free parameters. A similar improvement was found for the F814 data. A caveat here is that these face values should be taken with caution, given that additional systematic uncertainty would be present in both data and model and mitigate the difference in the chi-square value. We lastly comment that 
the half-light radii constrained by the single-S\'ersic model in the two bands differ by almost one order of magnitude, while such a difference is not seen for the double-S\'ersic model. 
This perhaps suggests that the double-S\'ersic model is preferable. 
However, to draw a robust conclusion on whether there are really two distinct S\'ersic components or not, for example, spatially resolved spectroscopy will be helpful, as it might be able to disentangle the  components kinematically.

\begin{figure*}
  \centering
  \includegraphics[scale=0.2]{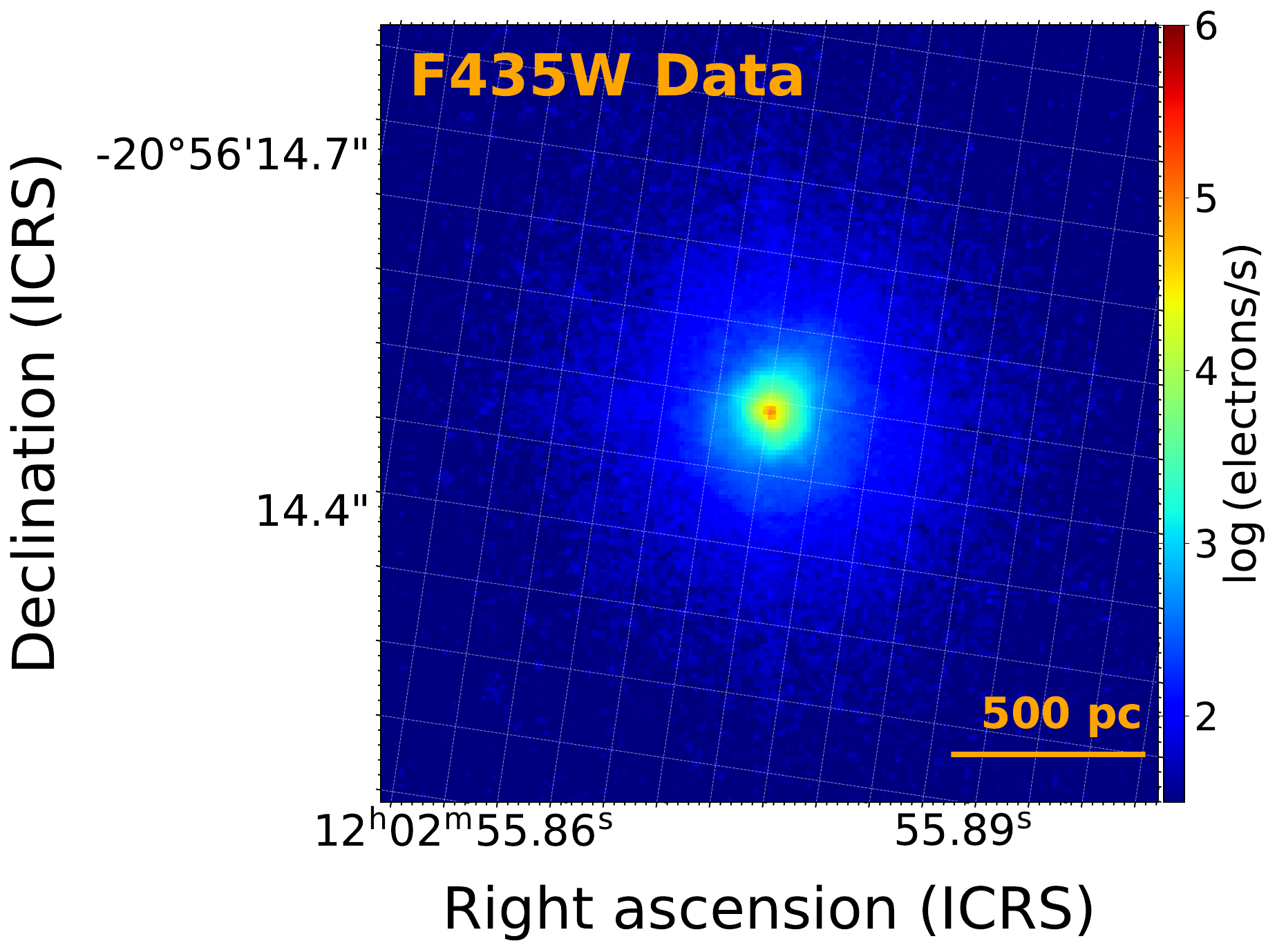}
  \includegraphics[scale=0.2]{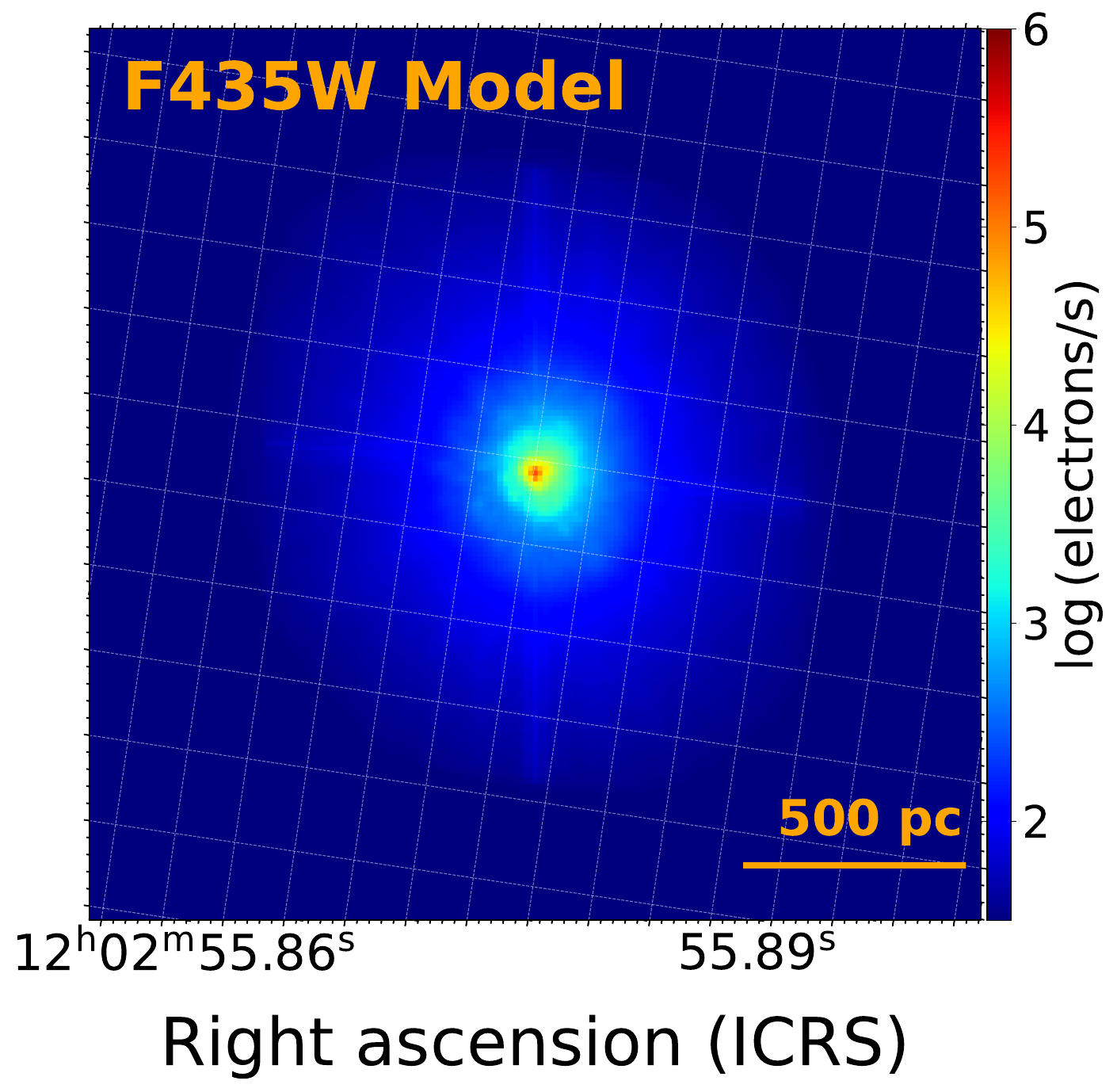} 
  \includegraphics[scale=0.2]{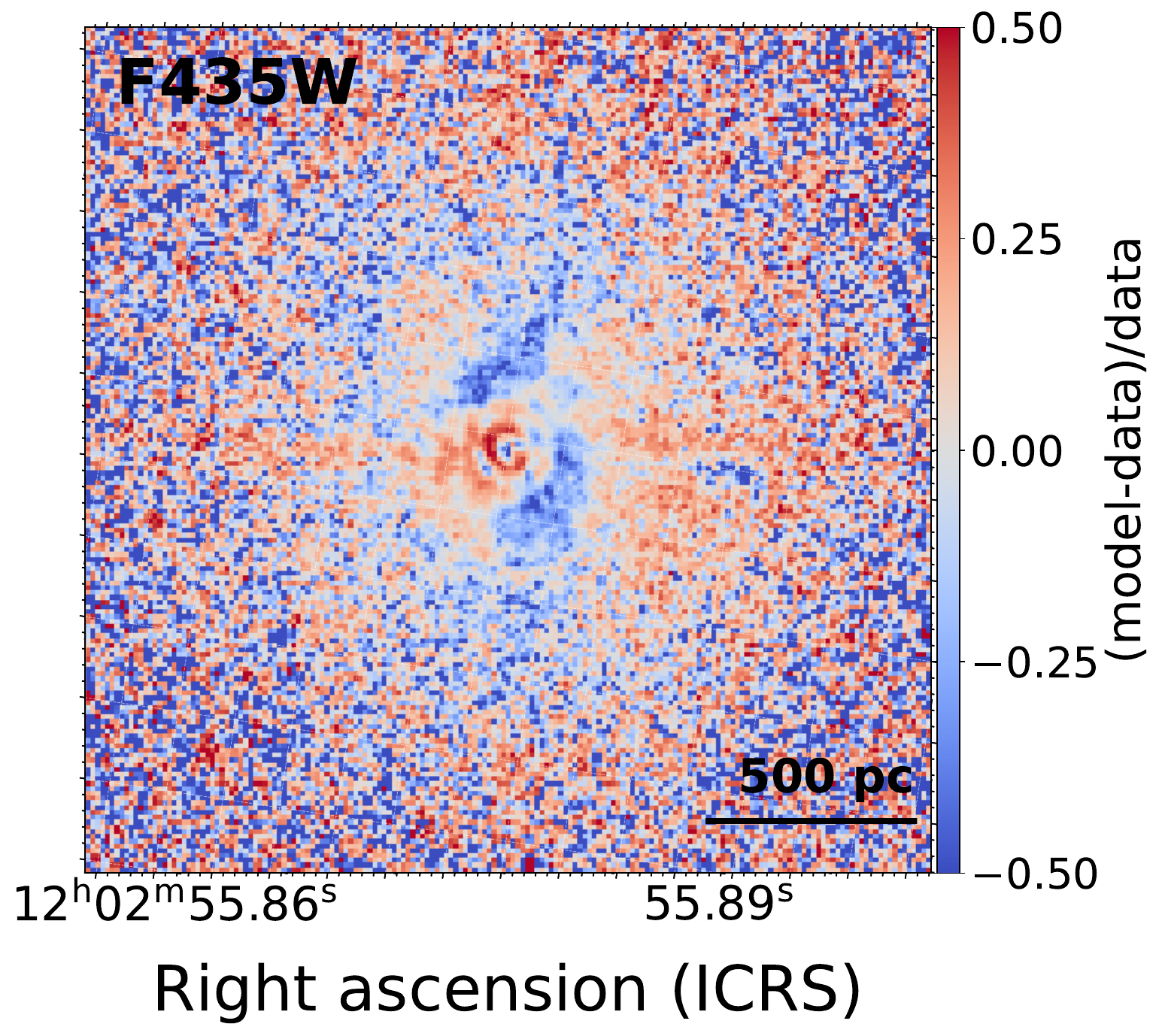} 
  \includegraphics[scale=0.2]{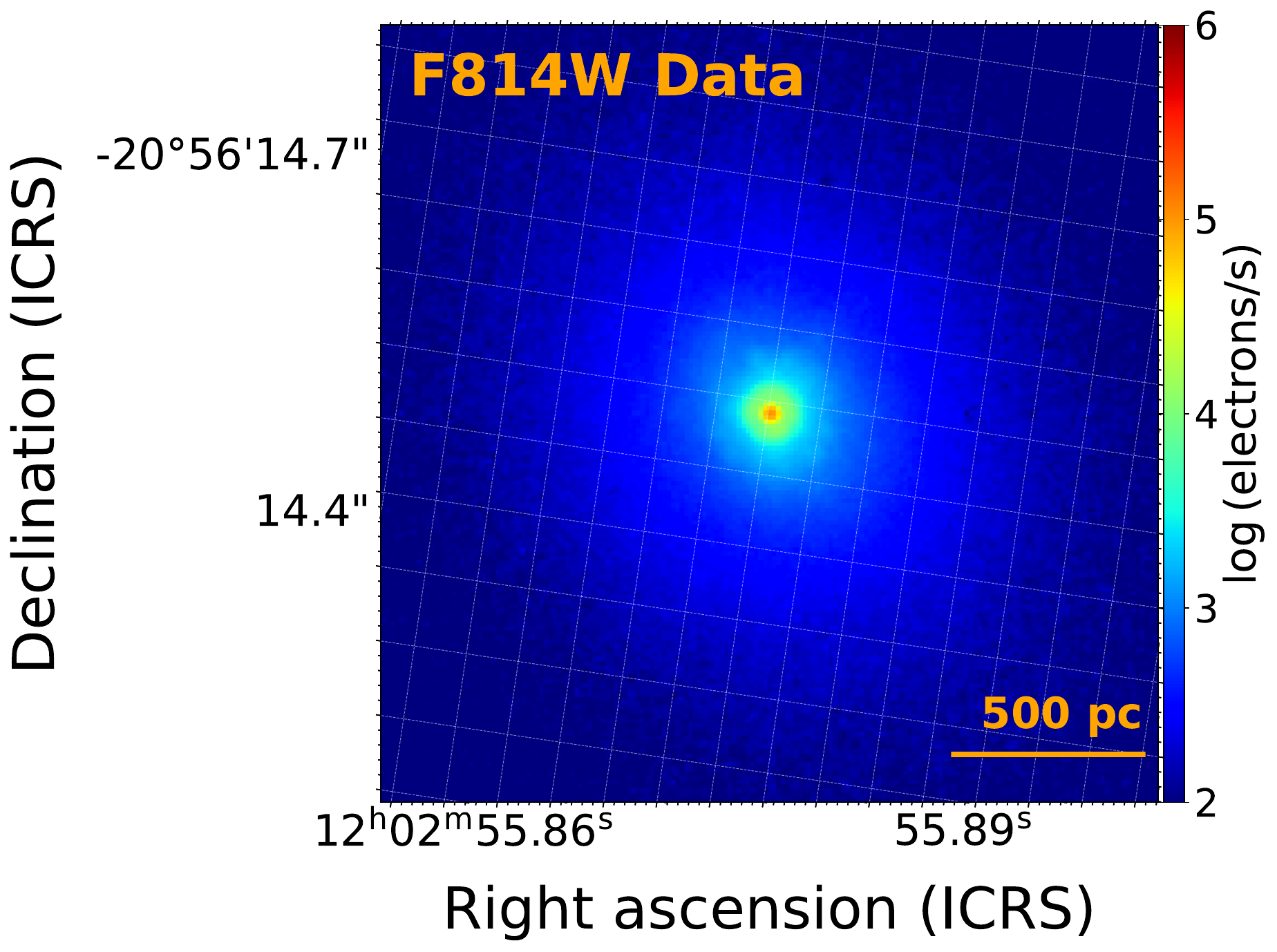} 
  \includegraphics[scale=0.2]{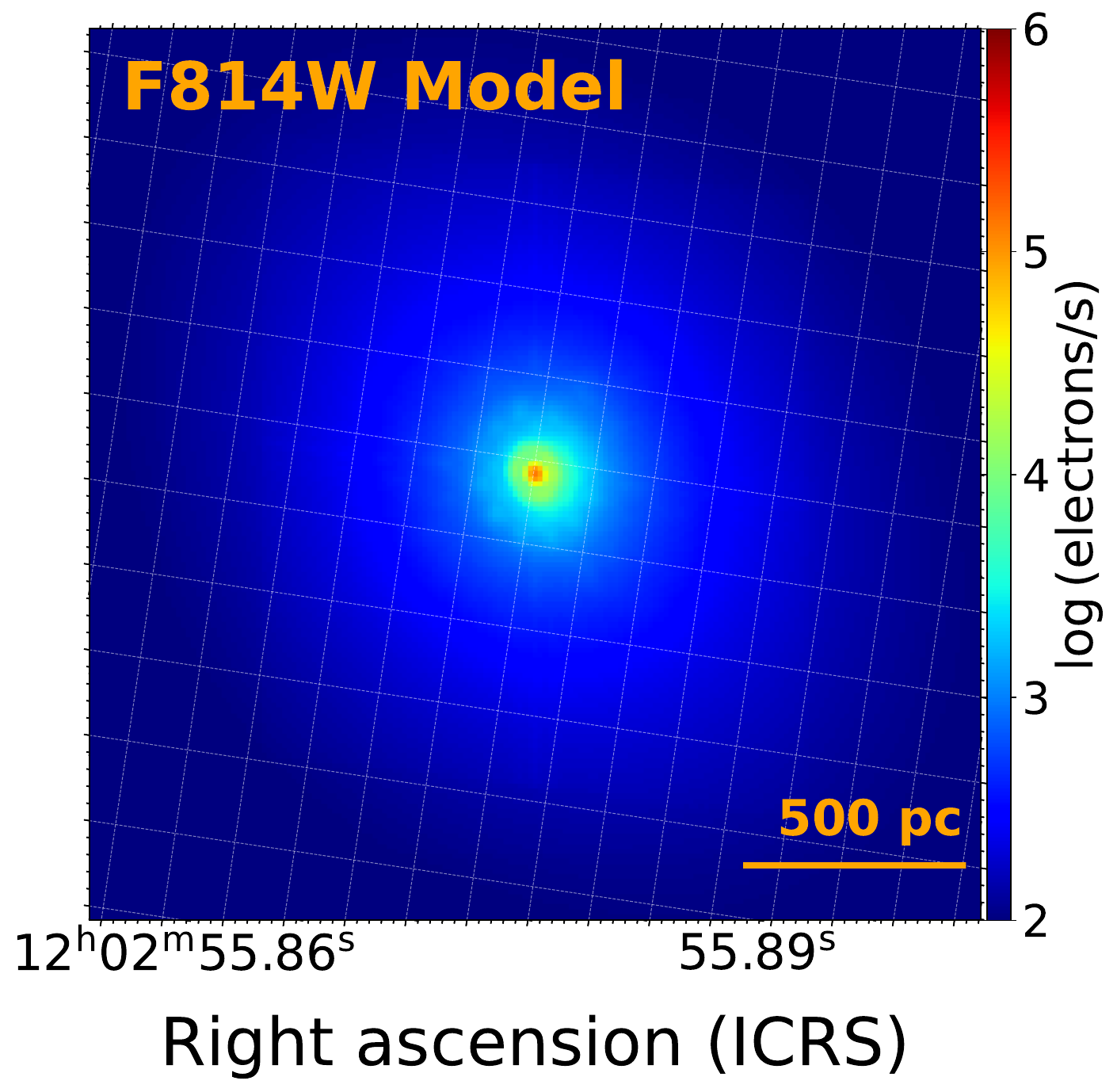} 
  \includegraphics[scale=0.2]{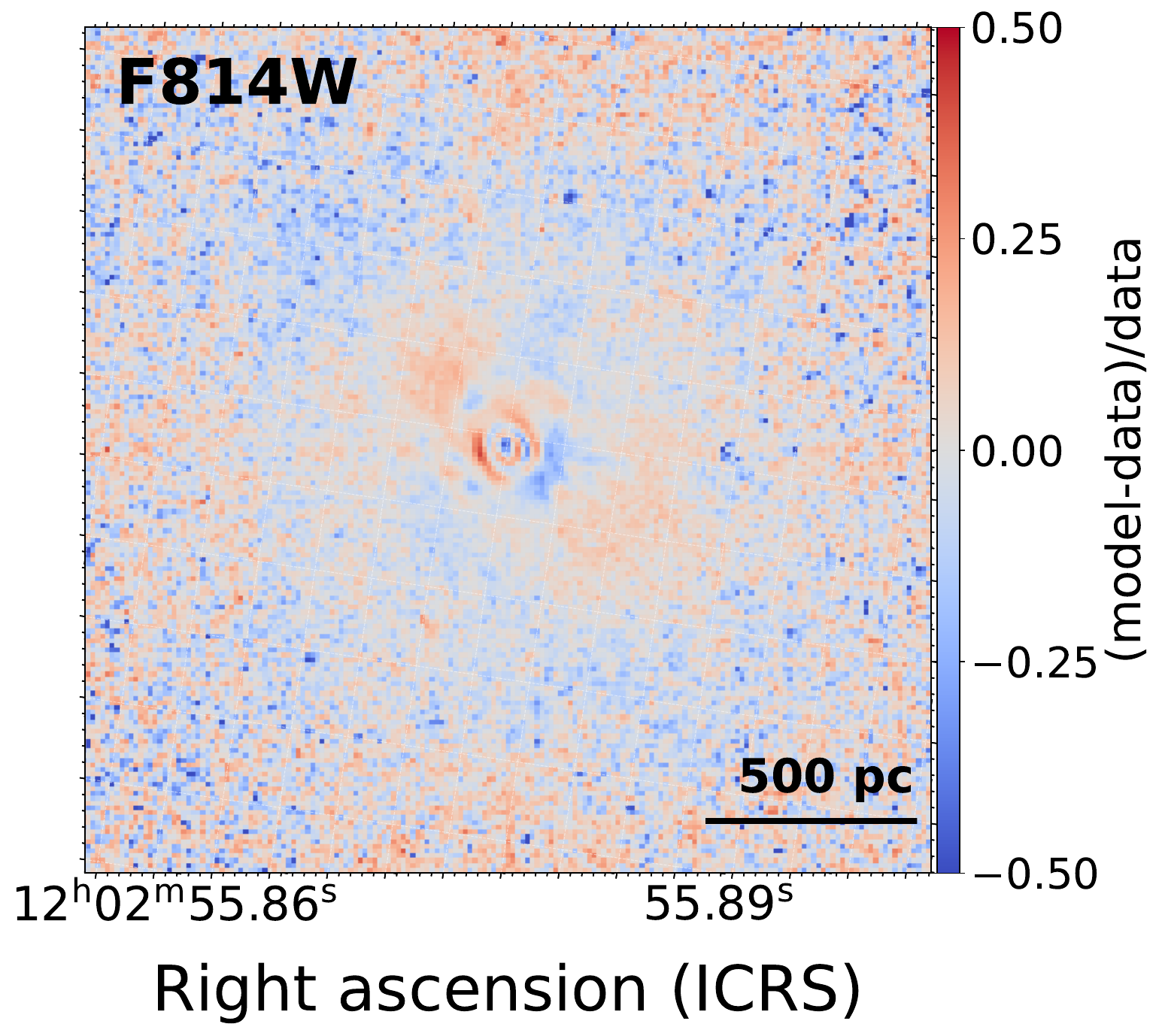}   
  \caption{
  Upper panels: From left to right, observed image, fitted model, and residual (difference divided by data counts) image in the F435W band. 
  Lower panels: Same as the upper panels, but for the F814W band. 
  }\label{fig:galfit2} 
\end{figure*}

In this paper, we preferentially adopt the results obtained by the double-S\'ersic model for discussions by considering its better description of the data and the similarity of the constrained parameters between the two bands. We, however, do not exclude the single-S\'ersic model. To show more on the good fits of the double-S\'ersic model, Figure~\ref{fig:galfit2} displays observed images, models, and their differences normalized by data counts. 
In addition, a result to be remarked on the double-S\'ersic model is 
that the S\'ersic profile with $n_{\rm S} = 4$ may represent the ``classical" bulge; the bulge-to-total flux ratios ($B/T$) in both bands are $\approx 0.6$ under the assumption that the profile with $n_{\rm S} = 4$ is a bulge component, and the values fall in an expected range for the classical bulge \citep{Gao22}.

The overall radial profiles seem to be well reproduced by the double-S\'ersic model as well as the single-S\'ersic one, as shown so far, but residuals, or (model -- data)/data, of $\sim 0.02$ are present at inner several pixels in both models (Figure~\ref{fig:galfit}). 
To quantitatively examine whether the residuals may be explained by systematic uncertainty in the PSF modeling, we analyzed \textit{HST}/HRC F435W and F814W data of GD 153, which is a white dwarf and can be assumed to be a point source. 
The data we analyzed were taken in 2004 February 15 and 2005 May 19, within which the \textit{HST} data of POX\,52 were obtained. 
In the same procedures as adopted for POX\,52, we reprocessed the \textit{HST} data and fitted the images with a PSF model and a sky-background model. We then found that, in both data, residuals of $\approx$ 0.03 remain, particularly in the inner pixels ($\lesssim$ 10 pix). Such residuals may be due to systematic uncertainties in the PSFs, and would explain the residuals found for POX 52.

Last but not least, we investigated what fitting results can be obtained for different assumptions on the fixed AGN magnitudes. 
Indeed, the AGN flux estimated by the SED fit is likely to be different from that during the \textit{HST} observations, due to intrinsic time variation. 
Possible variability can be inferred from the structure function (a measure of the ensemble rms magnitude difference as a function of the time difference between observations). \cite{Bal20} calculated the values of low-mass AGNs to be $0.2$ at most. Thus, we fitted each \textit{HST} image by changing the magnitude of the AGN component by $\pm$0.2 mag. 
As a result, for the single-S\'ersic model, the S\'ersic index was estimated to be $\gtrsim 4$ in both images. 
Also, in the case of the double-S\'ersic model, 
the $B/T$ ratios were estimated to be within 0.5--0.6, consistent with the above-derived values of $\approx$ 0.6. 
In summary, the impact is confirmed to be minor for our discussions.

\section{Estimate of the BH mass}\label{sec:bh}

We describe here our estimate of the BH mass of POX\,52, since it is important for explaining our modeling of the X-ray spectra. 
We estimated the BH mass based on the so-called single-epoch method \cite[e.g.,][]{Ves02,Ves06,Gre05}, for which an AGN luminosity and the width of a broad emission line are necessary. For the luminosity, we used the AGN luminosity at 5100\AA\, of \lopt =  1.1$\times10^{42}$ erg s$^{-1}$, which was constrained from the SED analysis and was confirmed to fit the surface brightnesses reasonably (Sections~\ref{sec:sed} and \ref{sec:galfit}). The uncertainty derived by CIGALE, as well as that inferred from the \textit{HST} analysis, is $\sim$ a few percentages. 
The broad emission line we adopted was H$\beta$, and, from \cite{Bar04}, we took the full width at half maximum of $v_{\rm FWHM} =$ 760$\pm$30 km s$^{-1}$. 
We substituted the two values to a calibrated equation used in the previous study of \cite{Tho08}, which referred to \citealt{Ben06} and \citealt{Onk04}:
\begin{eqnarray}
    M_{\rm BH} = 1.05\times10^7(\lambda L_{\lambda,5100}/10^{44} {\rm erg\,s}^{-1})^{0.518} \\ \nonumber
    \times (v_{\rm FWHM}/10^3 {\rm km\,s}^{-1})^2 M_\odot.
\end{eqnarray}
The mass was then estimated to be $\log (M_{\rm BH}/M_\odot) \approx 5.8$. 
As its error, we consider 0.5 dex or a factor of three, which 
is a canonical one and is dominated by systematic uncertainty 
\cite[e.g.,][]{Ves06,Ric22}.
Uncertainty should be present due to the measurements of the AGN luminosity and the line width in the different epochs. However, this would be negligible compared to 0.5\,dex, given a possibly small structure function of $< 0.2$ for low-mass AGNs \citep{Bal20}. 
Our estimated mass is slightly higher than the past one of $\log(M_{\rm BH}/M_\odot) \approx 5.5$--5.6 by \cite{Tho08}. This is because the AGN luminosity we used is 2--3 times higher than the past one of \cite{Tho08}, whose estimate was also based on a fit to the same F435W data. 
We rely on our estimate because, as described above, our AGN-luminosity measurement would be more reliable given that it fits the SED and also \textit{HST} images. An additional reason is because the newly estimated mass is more favorable for interpreting X-ray results, as described in the next section. 
Finally, we mention that \cite{Tho08} concluded that they could underestimate the luminosity. They found that the $B-I$ color 
of the galaxy components (i.e., two S\'ersic profiles) that they fitted to the \textit{HST} data was bluer than those of galaxies appropriate for comparison (e.g., Im galaxies) and interpreted the bluer color being due to 
the contamination of the AGN emission to the
S\'ersic components. 
Here, as lessons learned, we suggest that the SED analysis plays an important role in disentangling the AGN and host-galaxy components in the image fit. 
For further discussion, we adopt our mass estimate (i.e., $\log (M_{\rm BH}/M_\odot) \approx 5.8\pm0.5$) as our fiducial value. 

\section{X-ray Spectral and Timing Analysis}\label{sec:xspec}

To facilitate understanding of our X-ray analysis, we describe its flow here. 
We first jointly fitted the spectra obtained by the quasi-simultaneous \textit{NuSTAR} and \textit{XMM-Newton} observations (Section~\ref{sec:xspec_simple}). The spectra were averaged over the exposure times. 
The observation period of \textit{XMM-Newton} covers only part of the longer \textit{NuSTAR} observation (the right panel of Figure~\ref{fig:xlc}), but the flux measured by \textit{NuSTAR} for the partial period and the one for the whole period are similar. Thus, the spectral variation between the periods with and without \textit{XMM-Newton} could be neglected. 
The energy bands for \textit{XMM-Newton} and \textit{NuSTAR} we considered were 0.2--12 keV and 3--30\,keV, respectively. Events above 30\,keV were dominated by the background emission, and the spectra in that range were not considered. 
Since the spectral analysis alone was not sufficient to determine the final spectral model, we also discuss the X-ray variability during the course of our spectral analysis (Section~\ref{sec:xray_var}). 
Finally, we proceeded to the determination of the most plausible model (Section~\ref{sec:xspec_final}). After this analysis, we also discuss whether the time variability of the spectrum and the archival spectra can be readily explained by the model (Section~\ref{sec:xspec_var} and \ref{sec:xspec_past}).

\subsection{Simple Model Fitting}\label{sec:xspec_simple}

\begin{figure*}
  \centering
  \includegraphics[scale=0.5]{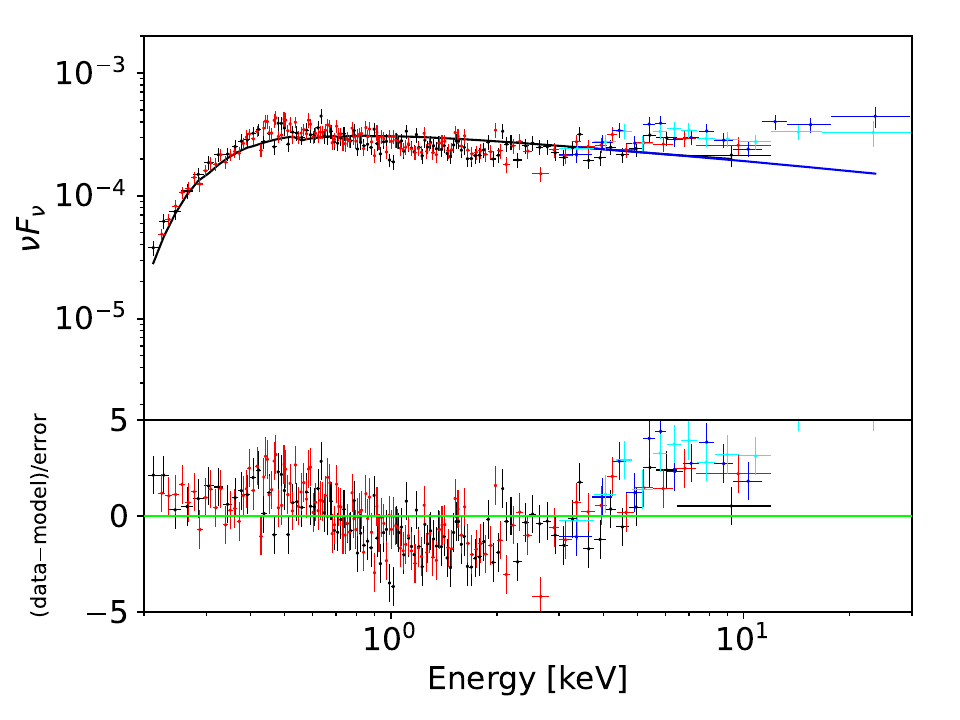}
  \includegraphics[scale=0.5]{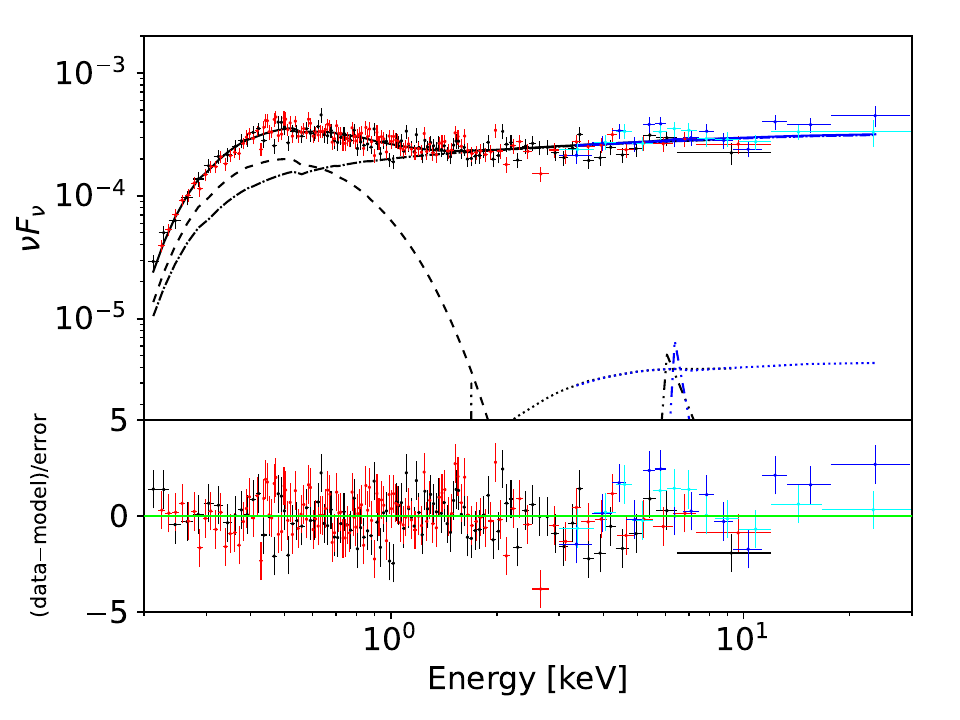} \\ 
  \includegraphics[scale=0.5]{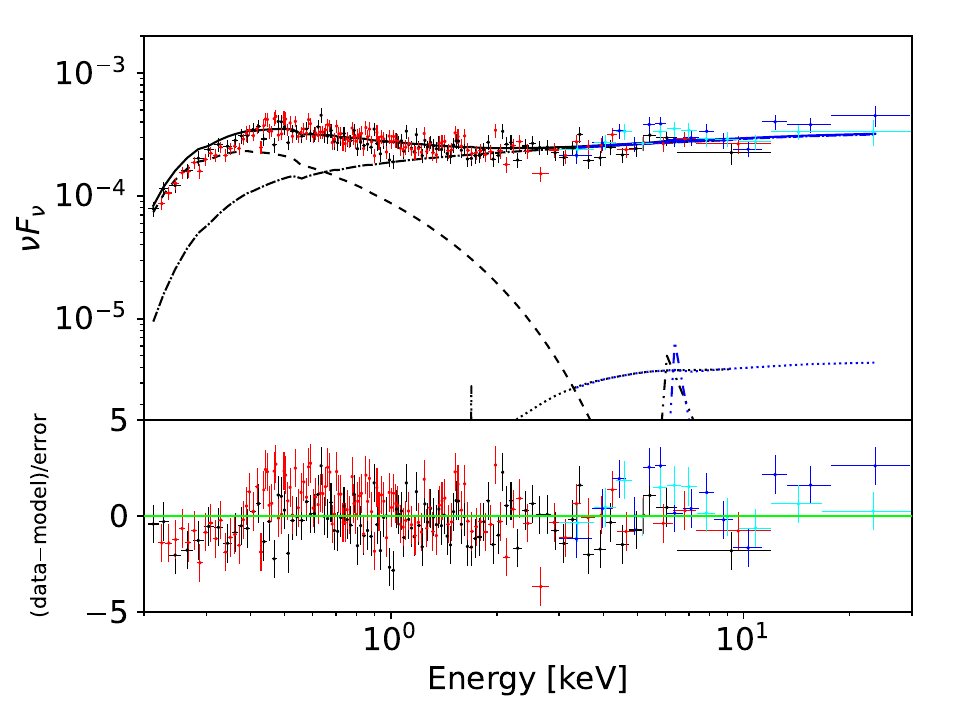} 
  \includegraphics[scale=0.5]{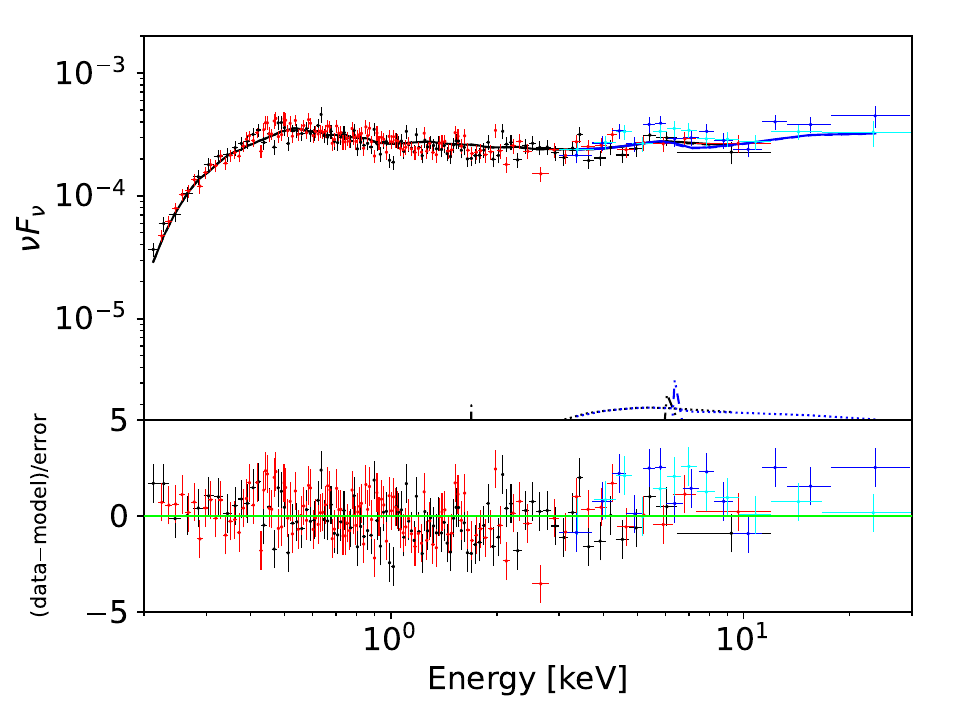} \\ 
  \includegraphics[scale=0.5]{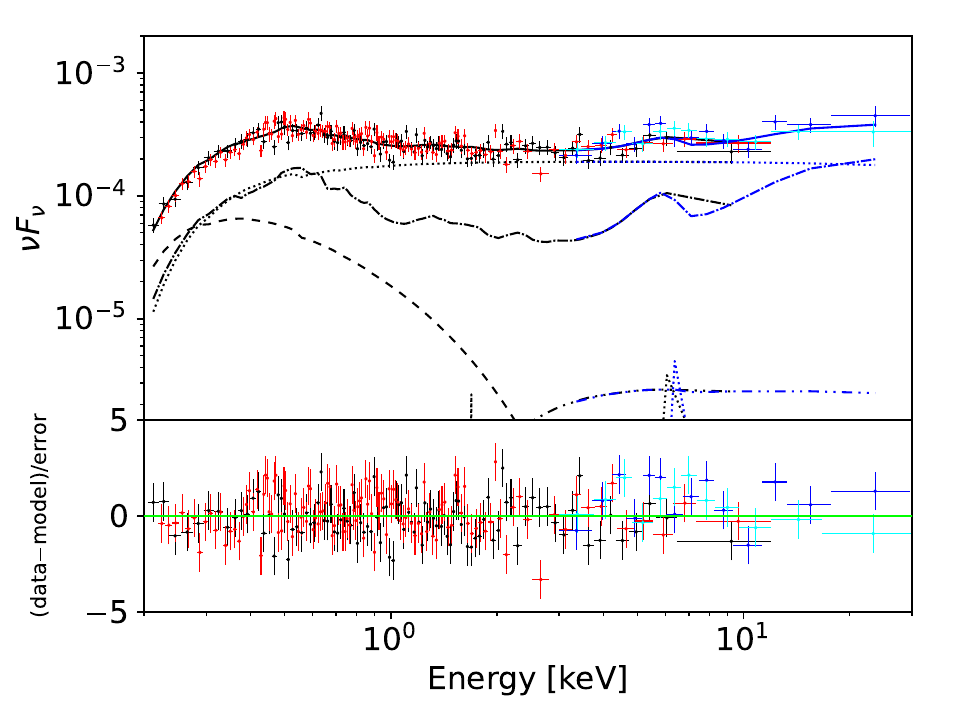} 
  \caption{
    Top-left: \textit{XMM-Newton} (black and red crosses) and \textit{NuSTAR} (blue and cyan crosses) 
    spectra fitted by an absorbed power-law model ($C$/d.o.f. = 2914.8/2782). Residuals are shown below the spectra. 
    Top-right: Improved fit by adding disk-blackbody (\texttt{diskbb}) and torus components ($C$/d.o.f. = 2510.4/2780), appearing particularly in the soft ($<$ 2\,keV) and hard band ($> 2$\,keV), respectively.  
    Middle-left: Fitting result by replacing the disk-blackbody component with the 
    warm corona model (\texttt{thcomp*diskbb}; $C$/d.o.f. = 2578.3/2778). 
    Middle-right: Spectra fitted by replacing the warm corona model with 
    a disk-reflection component (\texttt{relxilllpCp}; $C$/d.o.f. = 2532.0/2779). The primary power-law emission and reflected emission are shown as a combined component. 
    Bottom: Fitted by our final model consisting of the warm corona, power-law, disk-reflection, and torus-reflection components ($C$/d.o.f. = 2489.0/2775).  
  }\label{fig:xspec} 
\end{figure*}

As a baseline model, we considered the absorption by our galaxy and the primary X-ray power-law component with a high-energy cutoff originating in a hot corona around the IMBH; i.e., \texttt{tbabs*zhighect*zpowerlw} in the XSPEC terminology. 
The galactic absorption was fixed to $N_{\rm H}^\mathrm{Gal}\,=3.9\times 10^{20}$ cm$^{-2}$, derived using the HEADAS \texttt{nh} task \citep{Kal05}. 
The photon index and normalization of the power-law component were left as free parameters, while the cutoff was fixed at 200\,keV, typical for nearby AGNs \citep[e.g.,][]{Ric18cut}. 
By adopting the $C$-statistic method \citep{Cas79}, 
we fitted the model and found that the baseline model left clear excesses in the soft ($<$ 1\,keV) and hard ($>$ 10\,keV) bands, as shown in the top-left panel of Figure~\ref{fig:xspec}. Here, the resultant $C$ value is 2914.8 for the degrees of freedom (d.o.f.) of 2782.

To better reproduce the soft-band spectral feature, we added \texttt{diskbb} \citep[][]{Mit84}, often used to phenomenologically reproduce similar soft excesses of Seyfert galaxies \citep[e.g.,][]{Mid20,Jia20,Xu21}. The model computes the spectrum from a standard accretion disk for a given temperature at the innermost radius. 
For the hard band excess, we incorporated torus emission. The torus absorbs and scatters X-rays, producing a hump around 30\,keV, and a fraction of the absorbed X-rays result in fluorescent emission lines (e.g., the Fe line at 6.4\,keV). 
The torus emission was, indeed, suggested by the result of the SED analysis. 
It is natural to consider the same torus geometry used in the SED analysis; it is the SKIRTOR torus model, which considers a two-layer structure
\citep[][]{Sta16}, but there was no X-ray model that was built by considering the geometry. Therefore, we decided to alternatively adopt the clumpy torus structure, whose models were available in both IR and X-ray band \citep{Miy19,Yam23}. We adopted the IR one of \cite{Nen08_I} \citep[see also][]{Nen08_II} and the X-ray one of \cite{Tan19}.
First, we investigated with what parameters the IR clumpy torus of \cite{Nen08_I} can reproduce a spectrum close to that of our best-fit SKIRTOR model by varying clumpy-torus parameters: 
the ratio of inner and outer radii ($R_{\rm clump}$), the optical depth for each cloud ($\tau_{\rm V, \rm clump}$), the number of clouds along the equatorial plane ($N_{\rm clump}$), the radial-density-profile index ($q_{\rm clump}$), the angular width ($\sigma_{\rm clump}$), and the inclination angle ($\theta_{\rm inc}$). As a result, we found that the SKIRTOR-like spectrum was able to be reproduced with $R_{\rm clump} =$ 20, $\tau_{\rm V, \rm clump} =$ 40, $N_{\rm clump} = 1$, $q_{\rm clump} =$ 0.5, $\sigma_{\rm clump} =$ 15\arcdeg, and $\theta_{\rm inc} =$ 20\arcdeg. 
A comparison between the SKIRTOR and clumpy torus models was also discussed by \cite{Yam23}, and a similar conclusion was drawn by the authors. 
We note that we relied on the SKIRTOR model in the SED fitting, as it has been generally used in CIGALE and makes it easy to compare our result with other ones. 
The IR-torus geometrical parameters were then used to define our X-ray torus model, having 
hydrogen column density along the equatorial plane ($N^{\rm eqn}_{\rm H}$), torus angular width ($\sigma_{\rm clump}$), and inclination angle ($\theta_{\rm inc}$) as parameters. 
The column density was derived by converting the visual extinction of each clump to a column density under the assumption of a Galactic $N_{\rm H}$-to-$A_{\rm V}$ ratio \citep{Dra03} and by substituting the column density and also the number of clumps to Equation 5 of \cite{Tan19}. 
The equatorial column density was estimated to be $N^{\rm eqn}_{\rm H} \sim 10^{23}$ cm$^{-2}$. 
The two other parameters were simply fixed to the equivalent values of the IR torus model (i.e., $\theta_{\rm inc}$ and $\sigma_{\rm clump}$). 
As we set the incident X-ray spectrum to the primary X-ray one, no free parameters were left for the X-ray torus model. 
By adding these soft-excess and torus models, we obtained a much better fit with the $C =$ 2510.4 (i.e., $\Delta C \approx -400$) for the additional two free parameters (d.o.f. = 2780). 
The best-fit result shown in the top-right panel of Figure~\ref{fig:xspec} indicates that the soft excess is well reproduced by the \texttt{diskbb} model. On the other hand, the power-law emission becomes dominant in the X-ray band above $\approx$ 3\,keV rather than the torus emission, while the weak torus emission is consistent with no clear narrow Fe-K$\alpha$ emission at 6.4 keV. 

Although we obtained the good fit, we suggest that a different model is necessary, in particular, to describe the soft excess. 
The temperature of the \texttt{diskbb} component was constrained to be $\sim 160$\,eV. This is much higher than theoretically expected one of $\sim$ 20--30 eV for the standard disk model \cite[e.g., ][]{Sha73,Kub98,Mal22} where the inner edge extends down to the inner stable circular orbit (ISCO) for the spin parameter ($a_{\rm spin}$) of 0, $M_{\rm BH} \approx 6\times10^{5}\,M_\odot$, and the bolometric luminosity of $\sim 10^{43}$ erg s$^{-1}$. 
Thus, the \texttt{diskbb} component does not represent the inner part of the standard accretion disk, as established for basically all AGNs. This is also indicated from the normalization of the \texttt{diskbb} component; the normalization was constrained to be $\approx$ 65, and the corresponding inner radius of the disk was estimated to be $\approx 8\times10^9$ cm following \cite{Kub98}, much smaller than expected from the ISCO of $\approx 5\times10^{11}$ cm. 
We note that if we adopted \texttt{bbody} instead of \texttt{diskbb}, a temperature of $\approx$ 110 eV was obtained. Interestingly, this is consistent with a typical value measured for hard-X-ray selected type-1 AGNs, in spite of distinct difference in $M_{\rm BH}$ \citep{Ric17bass}. 



\subsection{Insight into Soft-excess Emission from the Flux Variation}\label{sec:xray_var}

\begin{figure}
  \centering
  \includegraphics[scale=0.5]{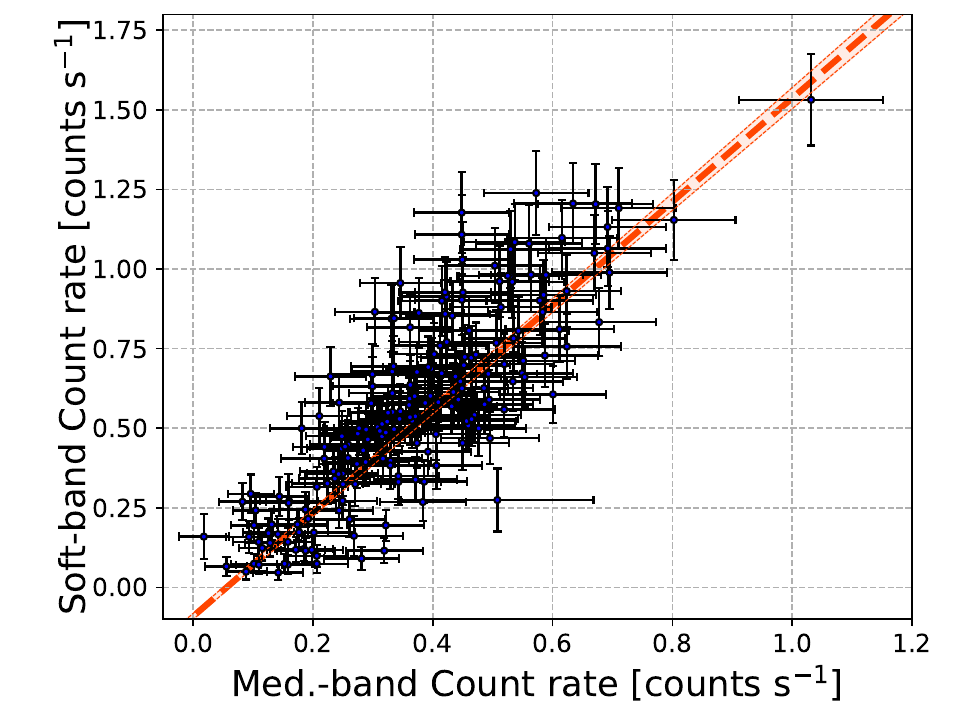}
  \includegraphics[scale=0.5]{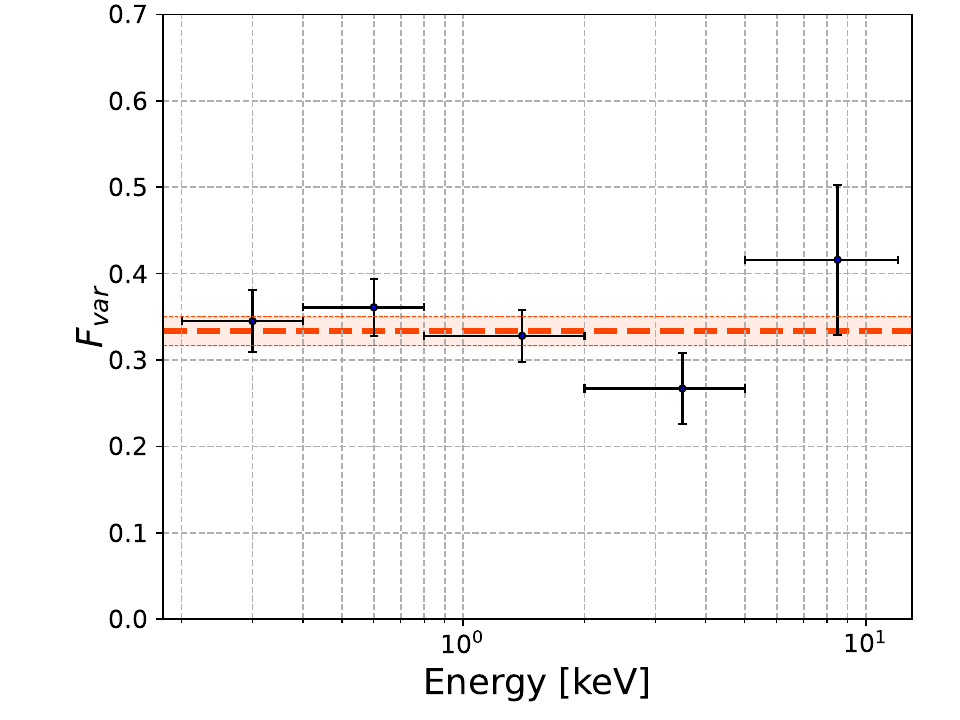}
  \caption{
  Top: Relation between the count rates in two bands of 
  0.2--1 keV and 1--12 keV, denoted as Soft-band 
  and Med.-band, respectively. Each count rate 
  is derived with a binsize of 100 sec, and all three detectors (MOS1, MOS2, and PN) are taken into consideration. 
  The linear regression line is drawn in orange together with the area corresponding to the 1$\sigma$ error. 
  Bottom: Fractional variability spectrum created with lightcurves with a binsize of 200 sec. This spectrum is consistent with being a constant model, as described by the orange dashed lines together with the 1$\sigma$ error. 
  }\label{fig:c2c} 
\end{figure}

To better understand what alternative component can be used for the soft excess, we focused on flux variation. 
Studying the variation, previous works have estimated the spatial scale of the soft-excess emitting region \citep[e.g.,][]{Noda13soft,Kam15,Lew22}, and we adopted the simple but powerful method of \cite{Noda11}. 
Their method is first to examine linear correlations of the count rate in a highly variable band \citep[e.g., $\sim$ 1--3\,keV;][]{Noda13n3516,Noda14} with those in different bands and then to reveal a component that follows the highly variable emission and the other stable component. 
Applying this method to five nearby ($z \lesssim 0.1$) Seyfert galaxies, \cite{Noda13soft} found a soft-excess component that did not vary synchronously with the primary power-law emission on a time scale of several hundred ks and 
concluded that it originated in a radius larger than several hundred gravitational radii.

Following \cite{Noda11}, we examined the correlation between 0.2--1 keV and 1--12 keV count rates for POX 52, as shown in Figure~\ref{fig:c2c}. The binsize was set to 100 sec so that almost all bins had at least a few counts. The result indicates that an offset for the soft-band count rate, which would have suggested the presence of a stable component, was not found. The same conclusion can be drawn even if we adopt similar energy bands (e.g., 0.2--1\,keV vs. 1--3\,keV) to those used by \cite{Noda11}. 
Thus, the soft excess could have varied on a time scale of 100 sec, or perhaps less, while following the power-law component. 
A consistent result was obtained by making a fractional variability spectrum with a similar binsize of 200 sec \citep[][]{Vau03}. 
This larger binsize was necessary, to keep photon counts in each of the divided energy channels.  
The spectrum in Figure~\ref{fig:c2c} was well reproduced with a constant model, consistent with the idea that fluxes across energy bands 
vary in a similar fashion within the time scale. 
These results suggest that the soft-excess emission originates in a region close to the IMBH, specifically, $r \lesssim 17\,R_{\rm g}$, or $\lesssim 3\,R_{\rm ISCO}$ if $a_{\rm spin} = 0$ and $\lesssim 13\,R_{\rm ISCO}$ if $a_{\rm spin} = 0.998$.

\subsection{Final Model}\label{sec:xspec_final}

Based on the above time-variability analyses, we tested two models. One was a warm corona model, and the other was an ionized-disk-reflection model. These have been often discussed as the origin of soft excesses seen in type-1 AGNs \cite[e.g.,][]{Pet13,Boi16,Pet18,Kub18}. 
In the warm corona scenario, such a secondary coronal component is usually presumed to cover a fraction of the accretion disk and, 
if the disk extends down closely to the central IMBH and the corona covers the inner region, rapid variability may be expected. 
Also, such a corona is optically thicker and cooler (i.e., $\tau_{\rm warm} \sim 10$ and $kT_{\rm e, warm} \sim 1$\,keV) than the hot corona responsible for the primary X-ray emission, and the spectrum of the warm corona model is generally harder than the disk thermal emission. Thus, it has the potential to reproduce the hot spectral feature revealed by fitting the \texttt{diskbb} model with $kT \sim 160$\,eV previously. 
On the other hand, the reflection model produces emission lines in the inner part of the ionized disk, and the blurring of the lines due to the relativistic effects results in the soft-excess-like spectrum \citep[e.g.,][]{Ros05,Cru06}. This is also expected to vary rapidly, given that it originates in the vicinity of the BH. 

We mention that, as a model to reproduce the soft-excess emission, a partial covering model is sometimes adopted, and we tested whether \texttt{zxipcf} can improve our final model presented in Section~\ref{sec:xray_combined}. 
We then confirmed that the partial covering model did not improve the fit significantly (i.e., $\Delta C \approx -4$ for $\Delta$ d.o.f. = 3). 

\subsubsection{Warm Corona Model}

We examined the validity of the warm corona model. The warm-corona Comptonization was modeled by \texttt{thcomp} with 
multi-temperature blackbody emission from the accretion disk as the seed-photon spectrum. The seed spectrum was modeled with \texttt{diskpbb}, which is different from \texttt{diskbb}, 
since it allows the user to vary the temperature distribution. In fact, the disk thermal emission has already been constrained in the SED analysis up to the UV region  (Section~\ref{sec:sed}), and some parameters of \texttt{diskpbb} were fixed so that it 
is consistent with our SED model. 
Specifically, by comparing the SED model and the \texttt{diskbpbb} spectrum in the optical/UV band ($\approx$ 0.1--1 $\mu$m), the normalization was fixed to 1.73$\times10^6$ and the power-law index that determines the temperature distribution was fixed to 0.637 (i.e., $p_{\rm disk}$ in $T \propto r^{-p_{\rm disk}}$). As a result, the free parameters of \texttt{thcomp*diskpbb} were the temperature at the inner edge of the disk, the optical thickness and temperature of the Comptonizing corona, and the covering factor of the corona. 

The result of our fit is shown in the middle-left panel of Figure~\ref{fig:xspec}. 
The temperature of the inner edge of the disk was constrained to be $\sim$ 30 eV, 
being in good agreement with the theoretically expected one for the disk reaching the ISCO. The $C$ value ($C$/d.o.f. = 2578.3/2778) is, however, worse than that obtained by adopting only the \texttt{diskbb} model. 
Also, the inconsistency between the model and the data is evident in the figure, particularly below 0.4\,keV. These facts would suggest that an alternative model is necessary to reproduce the data correctly.

\subsubsection{Disk Reflection Model}

We examined an ionized-disk-reflection model by using the \texttt{relxilllpCp} model \citep{Dau14,Gar14}. This model calculates reflected emission while considering relativistic effects. 
This was replaced with the power-law component, as the model returns both the primary power-law component and its reflected emission on the disk by executing a self-consistent calculation. 
Among the parameters to be set in the disk model, the inclination angle, the disk density, and the electron temperature of the corona, were fixed at 20\arcdeg, $10^{15}$ cm$^{-3}$, and 100 keV, respectively.  
The free parameters were the spin value, the inner radius, the photon index of the incident power-law spectrum, the ionization parameter, and the normalization. The other parameters were set to their default values. 
The resulting fit is shown in the middle-right panel of Figure~\ref{fig:xspec}. 
With fewer free parameters, this disk-reflection model better described 
the spectra ($C$/d.o.f. = 2532.0/2779) than the warm-corona model. 
However, the model appears to overestimate the data around 2--3\,keV and 
slightly underestimate the data in other bands. 
Also, this disk-reflection model would be incomplete, because it should be natural to expect the contribution of the tail of the thermal disk emission if the disk extends to the vicinity of the central IMBH and that of possibly associated Comptonization emission, like type-1 AGNs. Therefore, we further tested the model considering both Comptonization and disk-reflected emission, as detailed in the next subsection.

\subsubsection{Combined Model}\label{sec:xray_combined}

As a physically reasonable model for our broad-band X-ray data, we combined disk thermal emission, Comptonization, and disk-reflection and torus emission; i.e., \texttt{thcomp*diskpbb + relxilllpCp + [torus emission]}. 
While some of the parameters were fixed, similarly to what was done in the previous sections, we changed the disk density from $10^{15}$ cm$^{-3}$ to $10^{15.5}$ cm$^{-3}$. The main reason is clarified in the next paragraph. The best-fit model is shown in the bottom panel of Figure~\ref{fig:xspec}, where the power-law and reflected components are plotted separately for clarity. Reduced residuals are readily seen in the figure, and, in fact, the fit was significantly improved ($C$/d.o.f. = 2489.0/2775), compared with the disk-reflection model. 
The best-fit parameters obtained with this spectral model are summarized in Table~\ref{tab:xspec}. 
Here, to mitigate the degeneracy between the disk reflection 
\texttt{relxilllpCp} and the other disk emission (\texttt{thcomp*diskpbb}), we fixed the parameters of the latter emission and allowed only its normalization to vary synchronously with the disk emission in calculating errors. This is motivated by the result that the 0.2--1\,keV and 1--12\,keV emission varied in a similar fashion. 
The result suggests that the IMBH could spin rapidly ($a_{\rm spin} \approx 0.998_{-0.814}$, where the error corresponds to a 90\% confidence interval) and the inner radius of the reflecting disk is $3.2\,R_{\rm ISCO}$ (i.e., $\approx 4.0\,R_{\rm g} \sim 4\times10^{11}$\,cm).

We suggest that the composite model is reasonable as, in addition to the better statistical value,  no clear discrepancy exists between the constrained parameters.  
The thermal disk model (\texttt{diskpbb}) indicates the inner radius of the thermal disk to be $\sim  10^{12}$ cm ($\approx 7.2\,R_{\rm g}$) and the temperature at the disk inner-edge to be $\approx 24$\,eV. 
These values are consistent with those inferred 
by the disk-reflection model (\texttt{relxilllpCp}); 
the inner radius of the disk is $\sim 10^{11}$--$10^{12}$\,cm (the uncertainty in $M_{\rm BH}$ is taken into consideration), and the temperature expected at the radius is $\sim 30$\,eV. 
In addition, the ionization parameter indicated by the disk-reflection model is $\log \xi \approx 2.8$, and a consistent value can be derived by considering the disk density ($n_{\rm e} = 10^{15.5}$ cm$^{-3}$), irradiating X-ray luminosity ($\approx 10^{42}$ \ergs), and the inner radius of the disk ($\sim 10^{11}$--$10^{12}$ cm). 
Here, we note two things. If a smaller BH mass \citep[i.e., $M_{\rm BH}\approx 2$--4$\times10^{5}\,M_\odot$;][]{Tho08} is adopted, 
the inner-edge of the reflecting disk becomes smaller than 
inferred by the thermal disk emission, suggesting 
that the thermal disk cannot serve as the reflector. Thus, the larger mass is preferable. 
Also, if the disk density (i.e., $10^{15.5}$ cm$^{-3}$) is not changed from the default value of $10^{15}$ cm$^{-3}$, the calculated ionization can be higher than that suggested by the disk-reflection model. Thus, an increased density is favored. 
In summary, the model describes the broadband spectra without any difficulty in the parameters, and we treat it as our final model.

\begin{deluxetable}{cccc}
\tabletypesize{\footnotesize}
\tablecaption{Best-fit Parameters Obtained by Final Model \label{tab:xspec}}
\tablewidth{0pt} 
\startdata \vspace{-0.1cm} \\ 
   & Parameter & Best-fit  & Units \\ \hline \hline 
  \multicolumn{4}{c}{Galactic absorption}\\ \hline
  (1) & $N^{\rm Gal}_{\rm H}$ & $3.9\times10^{-2\,\dagger}$  & 10$^{22}$ cm$^{-2}$ \\ \hline \hline 
  \multicolumn{4}{c}{Comptonization of Multi-BB. emission (\texttt{thcomp*diskpbb})}\\ \hline
  (2) & $\tau_{\rm warm}$                   & 21$^\ddagger$        & ... \\
  (3) & $kT_{\rm e, warm}$  & 0.34$^\ddagger$    & keV \\
  (4) & $f_{\rm c, warm}$            & 0.10$^\ddagger$      & ... \\  
  (5) & $kT_{\rm e, disk}$  & 0.024$^\ddagger$   & keV \\  
  (6) & $p_{\rm disk}$       &  0.637$^{\dagger}$ & ... \\ \hline \hline
  \multicolumn{4}{c}{Disk reflection (\texttt{relxilllpCp})}\\ \hline  
  (7) & $\theta_{\rm inc}$     & 20$^{\dagger}$     & degrees \\
  (8) & $a_{\rm spin}$         &  $0.998_{-0.814}$   & ... \\
  (9) & $r_{\rm in}$           &  3.2$^{+1.5}_{-2.2}$         & ISCO \\  
  (10) & $\Gamma$               &  $1.98^{+0.04}_{-0.05}$ & ... \\
  (11) & $\log \xi_{\rm ion}$   &  $2.8\pm0.3$ &  ... \\  
  (12) & $n_{\rm disk}$         &  $10^{15.5\,\dagger}$ & cm$^{-3}$ \\    
  (13) & $kT^{\rm PL}_{\rm e}$  &  $100^{\dagger}$ & keV \\      
  (14) & Norm                   &  $6.0^{+0.6}_{-0.5}\times10^{-6}$ & ... \\ \hline \hline 
  \multicolumn{4}{c}{Torus reflection}\\ \hline    
  (15) & $N^{\rm Eqn}_{\rm H}$ &  $10^{23\,\dagger}$ & cm$^{-2}$ \\
  (16) & $\sigma_{\rm clump}$ &  $15^{\dagger}$ & degrees \\
  (17) & $\theta_{\rm inc}$ &  $20^{\dagger}$ & degrees \\ \hline \hline 
  \multicolumn{4}{c}{Flux and Luminosity}\\ \hline      
  (18)  & $F_{\rm 0.5-2}$ & 6.3 & 10$^{-13}$ erg cm$^{-2}$ s$^{-1}$\\   
  (19)  & $F_{\rm 2-10}$ & 6.7 & 10$^{-13}$ erg cm$^{-2}$ s$^{-1}$\\ 
  (20)  & $F_{\rm 10-30}$ & 6.4 & 10$^{-13}$ erg cm$^{-2}$ s$^{-1}$\\   
  (21) & $L^{\rm Comp}_{\rm 0.5-2}$ & 5.9 & 10$^{40}$ erg s$^{-1}$\\   
  (22) & $L^{\rm PL}_{\rm 2-10}$ & 4.9 & 10$^{41}$ erg s$^{-1}$\\ \hline \hline 
  \multicolumn{4}{c}{Statistical parameters}\\ \hline        
  (23) & $C$/d.o.f. & 2489.0/2775 & ... \\ 
\enddata
\tablecomments{
(1) Absorbing hydrogen column density due to our Galaxy in the sightline.  
(2,3,4) Optical depth, electron temperature, and covering factor of the warm corona. 
(5,6) Temperature at the inner edge of the disk and temperature-distribution index of the thermal disk component, providing seed photons to the warm corona. 
(7) Inclination angle from the polar axis of the disk. 
(8) Spin parameter. 
(9) Inner radius of the disk. 
(10) Photon index of the primary power-law emission. 
(11) Ionization parameter.
(12) Disk density. 
(13) Temperature of electrons responsible for the power-law emission. 
(14) Normalization. 
(15,16,17) Hydrogen column density along the equatorial plane, torus angular width, and inclination angle of the torus model. 
(18,19,20) Observed 0.5--2\,keV, 2--10\,keV, and 10--30\,keV fluxes. 
(21) Luminosity of the warm corona emission in the 0.5--2\,keV band. 
(22) Luminosity of the power-law, or hot corona, emission in the 2--10\,keV band. 
(23) $C$-statistic value versus d.o.f. 
Here, errors are quoted at the 90\% confidence level, following a convention in the X-ray analysis. 
$^\dagger$These parameters were fixed. 
$^\ddagger$ These parameters were fixed in calculating the other errors. See Section~\ref{sec:xray_combined} for more details. 
}
\end{deluxetable}

\begin{figure}
  \centering
  \includegraphics[scale=0.53]{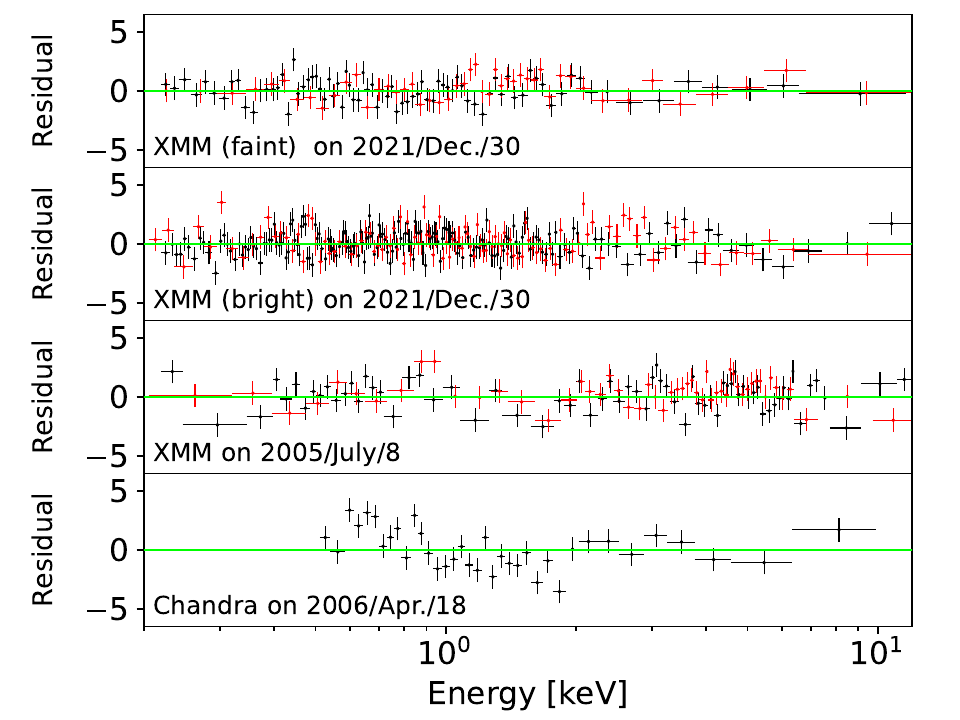}
  \caption{
  The first and second panels show residuals obtained when 
  we fitted the final model to the \textit{XMM-Newton} spectra in the fainter and brighter phases. 
  The third one shows residuals obtained by fitting the final model to the past \textit{XMM-Newton} spectra. 
  The last one shows residuals between the past \textit{Chandra} spectrum
  and our final fitted model. 
  }\label{fig:xspec_var_past} 
\end{figure}

\subsection{Spectral Variability}\label{sec:xspec_var}

The \textit{XMM-Newton} lightcurve shows strong variability (Figure~\ref{fig:xlc}), and thus, it is important to investigate whether the variation was the result of a change in the normalization or whether some other spectral parameters changed. To identify the reason, the observation period was divided into brighter and fainter phases, as indicated by blue and orange lines in Figure~\ref{fig:xlc}, and spectral fitting was performed in each phase. 
We eventually found that only by allowing the normalization of the entire model, except for the torus emission, to change freely, the spectra in the bright and faint periods can be fitted well without any clear residuals 
(Figure~\ref{fig:xspec_var_past}), or with $C$/d.o.f. = 1474.5/1767 and $C$/d.o.f. = 991.2/1186, respectively.

\subsection{Archival X-ray Spectra}\label{sec:xspec_past}

We finally assessed whether the final model could reproduce X-ray spectra obtained by previous X-ray observations carried out with \textit{XMM-Newton} on 2005 July 8 and {\it Chandra} on 2006 April 18. The observation IDs of the \textit{XMM-Newton} 
and \textit{Chandra} data are 0302420101 and 5736, respectively. 
These data were analyzed by \cite{Tho08}, and following their procedure, we extracted the source spectra from the \textit{XMM-Newton}/EPIC data in the 0.2--12 keV band and that from the \textit{Chandra}/ACIS data in the 0.5--7 keV band. 
To show that our model is plausible, we fitted the spectra with fewer free parameters. Specifically, for the \textit{Chandra} spectrum, we allowed only the entire normalization of the model to change freely. The fits resulted in $C$/d.o.f. = 421.2/356. While the statistical value is not as good as in the previous fits, clear residuals are not seen (Figure~\ref{fig:xspec_var_past}). 
Regarding the \textit{XMM-Newton} spectra, 
we included a partially covering ionized gas absorber model (\texttt{zxipcf}), whose presence in the spectra was already suggested
by \cite{Tho08}.
In the same way, as adopted for the \textit{Chandra} spectrum, only the overall normalization and parameters relevant to the \texttt{zxcipcf} model
were allowed to vary freely. We obtained a good fit ($C$/d.o.f. = 1466.429/1801; see also Figure~\ref{fig:xspec_var_past}). 
Between the spectra, including our new one, 
the X-ray flux varied; 
the 2--10\,keV flux observed by \textit{Chandra} was slightly higher by a factor of $\approx$ 1.1 than that during our new observations. 
In contrast, the flux during the past \textit{XMM-Newton} observation was $\approx$ 2.5 times lower. Even if corrected for the partial absorption, the flux was still 1.6 times lower.  
Thus, it was found that, in spite of the flux variation by a factor of $\sim 2$, our final model can reasonably fit the past spectra as well.



\section{Discussion}\label{sec:dis}

We have comprehensively revealed both host-galaxy and AGN properties based on the SED fit (Section~\ref{sec:sed}), the optical image fits (Section~\ref{sec:galfit}), and the X-ray spectral and timing analyses (Section~\ref{sec:xspec}). 
With the constraints obtained, we initially discuss the evolution of the system in Section~\ref{sec:evo}. Then, we discuss the AGN structure in detail in Section~\ref{sec:agn_str}. Finally, we summarize our overall AGN SED from the IR to the X-ray in Section~\ref{sec:sum_sed}. 

\subsection{Galaxy and Central IMBH Evolution}\label{sec:evo}

\begin{figure}
  \centering
  \includegraphics[scale=0.53]{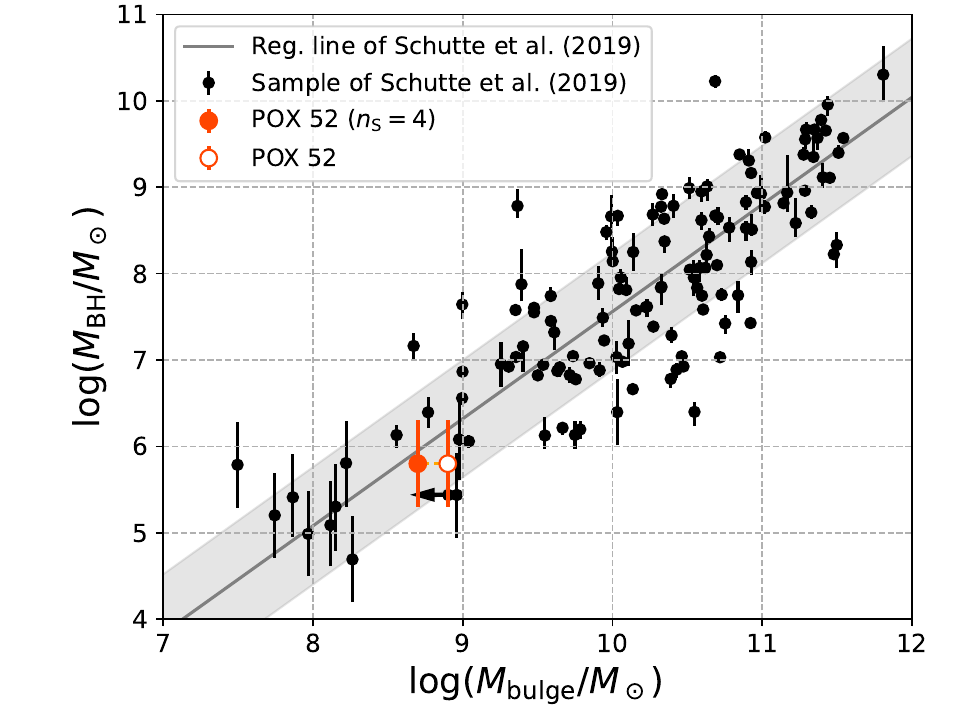}
  \caption{
    BH mass versus bulge mass. 
    Two orange dots are plotted for POX 52. 
    The filled one corresponds to the bulge mass that considers only the S\'ersic component with $n_{\rm S} = 4$ in the double-S\'ersic model (Section~\ref{sec:galfit}), while the unfilled one is the stellar mass derived by the SED analysis without any consideration of the imaging analysis.
  Galaxies investigated by \cite{Sch19} are 
  shown as black dots. Although POX 52 was included in their galaxy sample, the corresponding data point is removed here. The line and shaded area indicate
  a regression line and one standard deviation (0.68\,dex), derived by \cite{Sch19} for their sample. 
  }\label{fig:mbulge_mbh} 
\end{figure}

Our SED analysis suggests that the SF activity peaked possibly 
around $\sim$ 7.5 Gyrs ago, corresponding to $z \sim$ 1. 
As expected from this inference, the old stellar population ($>$ 10 Myr) with a mass of  $M^{\rm old}_{\rm star} \approx 8.3\times10^{8}$ $M_\odot$ dominates the entire stellar mass (i.e., $M_{\rm star} \approx M^{\rm old}_{\rm star}$), while the mass of young stars is $M^{\rm young}_{\rm star} \sim 10^6\,M_\odot$. 
Related to this, the \textit{HST} imaging analysis has given us an important insight into the host galaxy by finding that the classical bulge is the important structure. It is thus suggested that the bulge could have started to form at $z \sim 1$, and POX\,52 could have experienced a galaxy merger(s), given that the classical bulge is expected to be the result of the merger \cite[e.g.,][]{Naa03,Hop10bulge}. 
Currently at $z \approx 0.02$, the host galaxy of POX\,52 can be categorized as a star-forming galaxy given the SFR of 0.15 $M_\odot$ yr$^{-1}$ ($-$0.8 in a log scale) and the main sequence derived by \cite{Whi12}. The main sequence at $z < 0.5$ predicts an SFR for galaxies around the stellar mass of POX\,52 to be $10^{-0.4}$\,$M_\odot$\,yr$^{-1}$ with a scatter of 0.34\,dex, and this is consistent with the estimated SFR within $\approx$ 1$\sigma$.

The BH spin could provide important information on the growth history of the central BH, as suggested by numerical simulations \citep[e.g.,][]{Vol05,Ber08}. 
The BH spin can reach its maximum value of $\approx$ 0.998, if the mass
becomes about three times its initial value via prolonged accretion \citep{Tho74}. In contrast, if multiple BH mergers or stochastic accretion flows are the dominant way of growing for the BH, a lower spin value is expected on average. Although there is a very large uncertainty, the spin value was constrained to be $a_{\rm spin} = 0.998_{-0.814}$, which could suggest that the last growth episode was due to 
prolonged gas accretion. If this is true, the IMBH should have released the energy efficiently, perhaps affecting the host galaxy (i.e., through AGN feedback).  
Given the possible relation between the SF and AGN activity, 
AGN feedback could have been important around the peak of the SF ($z \sim$ 1). 
Here, we suspect that the current AGN activity would not be tightly linked to the presumed activity around $z \sim 1$, because an accretion episode since then at rates above the current one ($\lambda_{\rm Edd} \sim$ 0.3) would result in a mass of $\gg 1\times10^6\,M_\odot$ \citep{Sal64}. This is much larger than the current mass. Thus, it is likely that the current level of activity has been intermittent or has started recently. To reveal the actual mechanism that is feeding the central IMBH, it would be useful to study the properties of the cold molecular gas, which is the likely source of the AGN feeding \citep[e.g.,][]{Yam94,Mon11,Izu16,Sha20,Kaw21}.

To further investigate whether the host galaxy and the central IMBH of POX\,52 
have co-evolved while affecting each other, it is important to analyze the location of POX\,52 in the bulge/BH mass plot. 
We estimate the bulge mass to be $5.2$--$5.5\times10^8\,M_\odot$ by adopting the 
total stellar mass of $8.7\times10^8\,M_\odot$ and the bulge-to-total flux ratio of $B/T =$ 0.60--0.63 derivable from the analysis of the \textit{HST} data (Table~\ref{tab:galfit}). 
Our bulge mass is smaller than the previous ones of $\sim$1--2$\times10^{9}\,M_\odot$ \citep{Bar04,Tho08,Sch19}, and also our fiducial IMBH mass is $\log(M_{\rm BH}/M_\odot) \approx 5.8$, heavier than those estimated in the previous studies \citep{Bar04,Tho08}. 
As shown in Figure~\ref{fig:mbulge_mbh}, 
our results show that POX\,52 follows the $M_{\rm BH}$--$M_{\rm bulge}$ relation that was constrained by \cite{Sch19} down to $M_{\rm BH} \sim 10^5\,M_\odot$  and $M_{\rm bulge} \sim 10^8\,M_\odot$. 
This conclusion remains, even if no stellar decomposition is considered, or we adopt the single-S\'ersic model result. This is indicated in Figure~\ref{fig:mbulge_mbh} as well.
In addition, we find that with our estimates, 
POX\,52 almost perfectly follows the relation between $M_{\rm BH}$ and $M_{\rm star}$ determined for early-type galaxies down to $M_{\rm star} \sim 10^9\,M_\odot$ by \cite{Gre20}.

The fact that POX 52 follows the 
$M_{\rm BH}$--$M_{\rm bulge}$ and 
$M_{\rm BH}$--$M_{\rm star}$ relations could suggest that the central IMBH and host galaxy may have influenced each other.  
\cite{Kou21} investigated the $M_{\rm BH}$--$M_{\rm star}$ scaling relation using FABLE, simulating the cosmological growth of galaxies and BHs, with a particular focus on dwarf galaxies, and found that simulated BH masses tend to be smaller than expected by the observed relation. The authors suggested that the reason may be that the stellar feedback is too strong, which inhibits the growth of BHs 
\citep[see also][]{Tre18,Bel19,Kou19}. 
The prescription of the stellar feedback was based on that of \cite{Vog13}, which succeeded in explaining the observed properties of the galaxies (e.g., cosmic SFR history and stellar mass function). 
Later, \cite{Kou22} reported results for
different assumptions on the stellar feedback. 
Among their various simulation setups, they showed that BHs would have larger masses than expected from the observed $M_{\rm BH}$--$M_{\rm star}$ relation, especially in conditions where BH accretion can get close to the Eddington limit ($\lambda_{\rm Edd} \gtrsim 0.1$). 
This is due to the rapid growth of the BH and the suppression of SF by the AGN feedback. 
In addition, the authors also showed that 
less mass accretion rate onto a BH ($\lambda_{\rm Edd} \lesssim 0.01$) may be compatible with the $M_{\rm BH}$--$M_{\rm star}$ relation \citep[see Figures 2 and 8 of][]{Kou22}. 
Although the stellar mass range studied by \cite{Kou22} ($M_{\rm star} \sim 10^{5}$--$10^{8}\,M_\odot$) was smaller than the mass of POX\,52 ($M_{\rm star} \sim 10^{9}\,M_\odot$) by one order of magnitude, so we should be careful about relying on the results, POX\,52 and the observed relation could be reproduced if there was no extreme stellar feedback and moderate BH accretion, or AGN feedback, took place \citep[see also][]{Sha22}.

In summary, POX 52 would have experienced a galaxy merger(s), forming the classical bulge, and the IMBH has reached the current mass possibly through gas accretion at some point. Given that POX\,52 follows a $M_{\rm BH}$--$M_{\rm bulge}$ relation and the IMBH could have released accreting energy efficiently, AGN feedback could have taken place. 
The current level of AGN activity is, 
however, likely short-lived, and unrelated to the main channel of growth of the system.
The feeding mechanism at work now is unclear and, to reveal it, more data, especially cold molecular gas data, are necessary. 


%


\subsection{AGN Structure}\label{sec:agn_str}

X-ray data are important for revealing the structures of the AGN, especially in the vicinity of the central BH. By the quasi-simultaneous \textit{XMM-Newton} and \textit{NuSTAR} observations, we obtained the first broadband X-ray spectra of POX 52. Thus, the present study can provide a better understanding of the central structure through detailed X-ray data analysis than has been possible so far. 
The X-ray spectra were reproduced well with the multi-color temperature disk emission, Comptonization by the warm and hot coronae, and the reflected emission from the torus and the accretion disk (Table~\ref{tab:xspec}).

Soft-excess emission ($\lesssim$ 2\,keV) was observed, and could be reproduced by a disk-blackbody model with a characteristic temperature of $\sim$ 160 eV, similar to what were found for the target by \cite{Tho08} and other nearby AGNs \citep[e.g.,][]{Cru06,Bia09}. 
However, the soft excess does not represent the thermal disk emission,
because the temperature is much higher than expected from the BH mass and Eddington ratio of POX 52. 
Also, the estimated flux is much smaller than expected from an emitting region on a scale of the ISCO. 
Instead, while considering the discovery that soft (0.2--1\,keV) and hard (1--12\,keV) X-ray emission varies synchronously within a time scale of 100\,sec, corresponding to $r \lesssim 20\,R_{\rm g}$ (Section~\ref{sec:xray_var}), we suggest that the excess would consist of emission from a warm corona and disk-reflected radiation. 
The optical thickness and temperature of the warm corona were constrained to be $\approx 20$ and $\approx 0.3$\,keV, respectively. Similar thicknesses and temperatures were obtained for nearby Seyfert galaxies 
\cite[e.g.,][]{Don12,Jin12,Pet13}. Among various theoretical works, \cite{Pet18} proposed a warm corona model, succeeding in explaining the parameters measured for nearby AGNs ($z \lesssim 0.2$).
Our constraints can also be explained by their model, and, if we follow the model, the warm corona should be patchy (see Figure~1 of \citealt{Pet18}). 
This is indeed consistent with the low covering factor of our warm corona model (i.e., $\sim$ 0.1). 

The adopted thermal disk emission (\texttt{diskpbb}), providing seed photons for the warm corona, was fitted with a temperature of $\sim$ 20\,eV at the inner edge of the disk. 
This temperature is consistent with what is expected from the standard disk model 
with the BH mass and accretion rate of POX\,52 
\cite[e.g., ][]{Sha73,Kub98,Mal22} (see discussion in Section~\ref{sec:xray_combined}). 
The disk component was also constrained in the SED analysis, and its robustness was strengthened by the \textit{HST} imaging analysis. 
The slope of the component is, indeed, a valuable parameter that indicates the importance of advection in the accretion flow. 
The measured slope corresponds to the radial temperature distribution index of $p_{\rm disk} = $ 0.637, where 
$p_{\rm disk}$ is defined as $T \propto r^{-p_{\rm disk}}$. 
This is slightly smaller than a value expected by the standard disk model (i.e., $p_{\rm disk} =$ 0.75; \citealt{Sha73}). Given that 
the radial index of 0.5 is predicted for the slim disk, 
the radial advection may become important. Given that the AGN is accreting at a relatively high Eddington ratio ($\approx$ 0.3), the transition could be reasonable. 

The power-law component due to the hot corona was constrained, while the broad disk-reflection component was carefully considered. 
The measured photon index is rather soft, $\Gamma \approx$ 2, compared with the typical value of $\approx$ 1.8 for nearby hard-X-ray selected AGNs \citep[e.g.,][]{Ric17bass}, but this is consistent with what is expected from the (weak) positive correlation between the photon index and Eddington ratio, reported so far \cite[e.g.,][]{She06,She08,Kaw16b,Tra17}.
Because we have good constraints on both X-ray power-law and disk components, the relative strength of the X-ray luminosity to the bolometric one can also be estimated well. We here define the bolometric luminosity as their total luminosity in the energy range of 0.001--100 keV, as adopted by \cite{Vas09}. The estimated bolometric luminosity is $L_{\rm bol} = 2.2\times10^{43}$\,erg\,s$^{-1}$, where the disk (0.001--100\,keV) and X-ray power-law luminosities (0.1--100\,keV) are $2.0\times10^{43}$\,\ergs and $2.0\times10^{42}$\,\ergs, respectively. 
The Eddington ratio is thus $\lambda_{\rm Edd} \approx$ 0.3. Considering that 
the 2--10 keV luminosity of the power-law component ($L^{\rm PL}_{\rm 2-10} = 4.9\times10^{41}$\,\ergs), the bolometric correction factor is $\approx$ 45. This is consistent with what is expected from an 
Eddington-ratio dependent bolometric correction factor of  \cite{Vas09} \citep[see also][]{Vas07,Dur20}, rather than an expected value from a luminosity dependent one \cite[e.g., $\sim$ 10 from][]{Mar04}. 
This result suggests that POX 52 forms a similar disk-corona system, including the warm corona, to those typically used to explain the observational properties of Seyfert galaxies. 
Also, we suggest that even for low-mass AGNs with $M_{\rm BH} \sim 10^{5-6}\,M_\odot$ accreting at an Eddington ratio of $\approx 0.3$, a similar bolometric correction factor could be used.

The SED analysis, and especially the IRS spectrum, revealed that the emission from the dusty torus is very weak and its solid angle and optical thickness are small. The X-ray model assuming such a torus structure is consistent with the observed X-ray spectra. In particular, the weak Fe-K$\alpha$ emission line at 6.4\,keV supports a poorly developed torus. The dependence of the torus solid angle on AGN parameters has been much investigated, and its dependence on the Eddington ratio has been revealed \citep[][]{Ric17nat,Cam21}. 
Physically, it is proposed that as the Eddington ratio increases, the radiation pressure becomes stronger relative to gravity, and the surrounding dust and gas can be more easily blown out \citep{Fab08,Fab09}. 
The constrained angle of the torus from the equator is $\sigma_{\rm clump} = $ 15\arcdeg, and 
with the number of clumps along the equatorial plane ($N_{\rm clump} = 1$), the covering factor is estimated to be $< 0.1$ (see Equation 9 of \citealt{Nen08_I}). 
This value is expected from the decrease of the covering factor with Eddington ratio, although it is a bit smaller than a predicted range (i.e., $\sim$ 0.2--0.4 around $\log \lambda_{\rm Edd} \approx -0.5$). 
Thus, the radiation pressure could play a role in shaping the torus, but it would be important to 
consider a different factor as well. 

As is inferred from the small torus covering factor, 
the torus luminosity in the mid-IR (MIR) band is considerably lower than expected from the MIR-to-AGN luminosity correlation \citep[e.g.,][]{Gan09,Asm15,Ich17}. 
For example, if we adopt a $L_{\rm 12\,\mu m}$-to-$L_{\rm 2-10\,keV}$ relation derived for nearby type-1 AGNs in \cite{Asm15}, who 
measured $L_{\rm 12\,\mu m}$ at sub-arcsec resolutions, 
an expected MIR luminosity is $\log(L_{\rm 12\,\mu m}/{\rm erg\,s^{-1}}) = $ 42.1$\pm0.25$. However, our SED analysis indicated 
a considerably lower 12\,$\mu$m luminosity of $40.9$ (i.e., $\Delta \log L_{\rm 12\,\mu m} \sim 5\sigma$). 
Even if the possible dependence of the MIR(torus)-to-AGN luminosity ratio on the Eddington ratio is considered, there still seems to be a discrepancy. 
For example, the work of \cite{Tob21} 
(see their Figures 12 and 13) suggest $L^{\rm torus}_{\rm IR}/L_{\rm bol} \sim$ 0.1--0.4 at $\lambda_{\rm Edd} \approx$ 0.3, while for POX\,52 is $\sim 0.003$. 
Given that the observed trends are often interpreted as being due to an interplay between the central engine and torus, a different factor from it would need to be considered. 
One of the possible reasons might be 
the small amount of gas and dust in the galaxy, expected as 
POX 52 is a dwarf galaxy. 
Since it is difficult to discuss the 
MIR deficit furthermore based on the current data, it would be an idea to study POX\,52 in the future with far-IR and 
(sub)millimeter-wave observations to examine the cold dust and gas properties. 


In summary, the corona-disk system of POX\,52 is similar to those suggested for typical type-1 AGNs, and its nature as well as that of the torus, except for the considerable MIR deficit, are found to be consistent with what are expected from the Eddington-ratio dependent scenarios. 
In addition, we remark that the flatter radial temperature profile ($p_{\rm disk} < 0.75$) could suggest that the advection becomes important. 


\subsection{An SED Template of Rapidly Accreting IMBH} \label{sec:sum_sed}

\begin{figure*}
  \centering
  \includegraphics[scale=0.9]{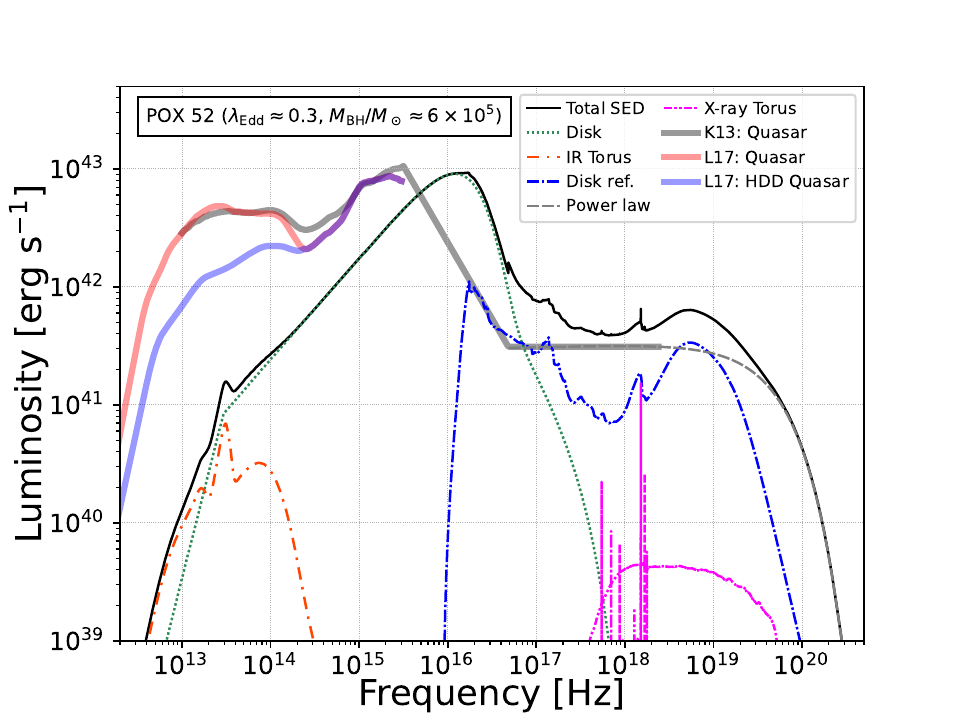}
  \caption{
  Model AGN SED of POX 52, composed of the thermal torus emission, the thermal and partially Comptonized disk emission, the primary X-ray power-law emission, and 
  the X-ray reflected emission from the disk and the torus. The legend indicates their corresponding lines and colors. 
  This SED is corrected for the dust-extinction and absorption. 
  For comparison, a quasar SED constructed by \cite{Kra13} is plotted by a thick gray line,
  and two SEDs from \cite{Lyu17} are also shown: normal-quasar SED in red and HDD-quasar SED in blue. 
  The SED of \cite{Kra13} is adjusted from the original one so that its X-ray power-law component fits that for POX 52. 
  On the other hand, the SEDs of \cite{Lyu17} are re-scaled so that they smoothly connect with the SED of \cite{Kra13} at $10^{15}$\,Hz. 
  }\label{fig:agnsed} 
\end{figure*}

Model SEDs of IMBHs accreting at high Eddington ratios ($\lambda_{\rm Edd} >$ 0.1) are important in discussing whether or not existing and future telescopes will be able to detect a similar population at high redshifts, or seed BHs of quasars found at $z > 6$ \citep[][]{Mar20,Gri20}. Detecting the seed BHs is one of the most important tasks toward a better understanding of how high-$z$  quasars have grown in a limited time \cite[e.g.,][]{Wu15,Ban18,Ono19}. 
The small BH mass and high Eddington ratio of the AGN in POX 52 resemble progenitors expected before the high-$z$ quasars; thus, the AGN SED of POX 52 we constrained could be useful for the discussion. 
We summarize the extinction- and absorption-corrected AGN SED of POX\,52 in Table~\ref{tab:agnsed}, and show it in Figure~\ref{fig:agnsed}. 

In Figure~\ref{fig:agnsed}, for comparison with more massive AGNs, we also show a quasar SED of \cite{Kra13}, re-scaled so that its X-ray power-law component matches that of POX 52. 
The authors  
compiled SEDs of luminous broad-line quasars at $z = $ 0.064--5.46, and we selected this as this covers the broadband wavelength from X-ray to IR. 
In addition, the other two SEDs from the study of \cite{Lyu17} are plotted. The authors studied IR SEDs of the PG quasars at $z \lesssim 0.5$, and classified them into three categories: normal quasar, warm-dust-deficit quasar, and hot-dust-deficit (HDD) quasar. 
In this order, the MIR emission becomes weaker with respect to the optical one. 
Among the three provided SEDs, we show the extreme two of the normal-quasar SED and the HDD-quasar SED. The latter is adopted, mainly because POX 52 shows weak MIR emission, as described previously.  
The Eddington ratios of the HDD quasars ($\lambda_{\rm Edd} \lesssim 0.1$) tend to be lower than those of the normal quasars ($\lambda_{\rm Edd} = 0.1$--1). Considering this fact, \cite{Lyu17} proposed an idea that at the lower Eddington ratios, the disk gets thicker and the ambient gas density is lowered, resulting in less efficient production of host dust emission (see the original paper for more details). An interesting argument is that the weak MIR emission of high-$z$ dust-free, or dust-poor, quasars \citep{Jia10,Hao10} can be explained by their idea. 

The comparison to the SED of \cite{Kra13} shows a distinct difference in the peak of the disk emission. This would be however explained by considering that the low-$M_{\rm BH}$ AGN can achieve a hotter temperature at the inner edge of the disk than more massive AGNs, or quasars. 
The comparison to the SEDs of \cite{Lyu17} also gives us an important insight. Particularly, even the HDD-quasar SED cannot reproduce the weak IR emission of POX 52.  Thus, a factor different from the interplay between the disk and torus that is considered in \cite{Lyu17} may be important, such as the small amount of gas and dust in the galaxy.

\begin{deluxetable*}{ccccccccccc}
\tablecaption{Data of the AGN SED and spectral components\label{tab:agnsed}} 
\tablewidth{0pt} 
\startdata \vspace{-0.1cm} \\ 
  Frequency  & Total & Accretion Disk & IR Torus & X-ray Power-law & X-ray Disk Reflection & X-ray Torus \\ \hline 
1.31e+17&7.73e+41&1.15e+41&NaN&3.09e+41&3.49e+41&3.76e+37\\
1.63e+17&6.08e+41&7.85e+40&NaN&3.10e+41&2.19e+41&5.67e+37\\
2.03e+17&5.21e+41&5.17e+40&NaN&3.11e+41&1.58e+41&9.95e+37\\
3.16e+17&4.49e+41&1.86e+40&NaN&3.13e+41&1.17e+41&4.63e+38\\
3.17e+17&4.49e+41&1.85e+40&NaN&3.13e+41&1.17e+41&4.67e+38\\
\enddata
\tablecomments{
Frequency is in units of Hz, and the others are in units of erg s$^{-1}$. 
The same data are used to plot the AGN SED and spectral components in Figure~\ref{fig:agnsed}. 
Luminosities below $10^{35}$ erg s$^{-1}$ are just denoted as NaN. 
We note that, due to this luminosity cut, the ``IR Torus'' values in this table are NaN, but finite values are listed in a frequency range of $\sim 10^{12}$--$10^{15}$ Hz. 
The entirety of this table can be obtained online. 
}
\end{deluxetable*}

\section{Summary}\label{sec:sum}

POX 52 is a nearby (93\,Mpc) dwarf galaxy hosting a rapidly growing low-mass AGN with $\log(M_{\rm BH}/M_\odot) \approx 5.8$ and $\log \lambda_{\rm Edd} \approx -0.5$. This is one of the best galaxies to reveal the host and AGN properties in low-mass systems and to discuss how such an object has evolved and how its nucleus is structured. 
For the discussion, we collected the multi-wavelength data from the X-ray to the radio. Particularly, the X-ray data were newly obtained by the quasi-simultaneous \textit{NuSTAR} and \textit{XMM-Newton} observations. 
Our analyses and results obtained with the broadband data are summarized as follows. 

\begin{itemize}
    \item We constructed the SED from UV to IR using the \textit{XMM-Newton}/OM, \textit{GALEX}, PanSTARRS, 2MASS, \textit{WISE}, and \textit{Spitzer}/IRS data, and fitted it via the CIGALE code (Section~\ref{sec:sed}). 
    The result (Table~\ref{tab:cigale}) suggests that the host-galaxy has grown with the peak of SF activity being at $z \sim 1$, and the stellar light is now dominated by old stars with an accumulated mass of $M^{\rm old}_{\rm star} \approx 8.3\times10^8\,M_\odot$. 
    Currently, at $z \approx 0.02$, the SFR is 
    0.15 $M_\odot$ yr$^{-1}$, and the galaxy can be categorized as a star-forming galaxy. 

    \item We decomposed the high-resolution optical \textit{HST} images around 4300\AA\, and 8100\AA\ into AGN and host-galaxy components using the GALFIT code (Section~\ref{sec:galfit} and Figure~\ref{fig:galfit}), while fixing the AGN flux densities to the values constrained in the SED analysis. 
    Then, the possible presence of a classical bulge was revealed. This was 
    supported by the moderate bulge-to-total flux ratio of $\approx$ 0.6. 
    Considering this ratio and the total mass, the bulge mass was estimated to be $M_{\rm bulge} \approx 5\times10^{8}\,M_\odot$. 
    Through this analysis, as a lesson learned, an unreasonable result (largely different AGN contributions in two adjacent bands) was obtained without fixing the AGN contributions, indicating the importance of simultaneous SED and image analyses. 

    \item We have estimated the BH mass to be  $\log(M_{\rm BH}/M_\odot) \approx 5.8\pm0.5$ by adopting the newly constrained optical AGN emission in the single-epoch method (Section~\ref{sec:bh}). 
    
    \item We fitted the broadband X-ray spectrum (0.2--30\,keV) obtained by the quasi-simultaneous \textit{NuSTAR} and \textit{XMM-Newton} observations. 
    While considering flux variation (Section~\ref{sec:xray_var}) and the disk and torus components constrained in the SED analysis (Section~\ref{sec:sed}), 
    we found a reasonable X-ray model, considering thermal and Comptonized disk emission, the primary X-ray power-law emission, and the reflected emission from the disk and torus (the bottom panel of Figure~\ref{fig:xspec}). 
    The BH spin parameter was roughly estimated to be $a_{\rm spin} = 0.998_{-0.814}$ at a 90\% confidence level (Table~\ref{tab:xspec}). This could suggest that most of the current mass was achieved by prolonged mass accretion. 

    \item We have discussed a possible scenario on the evolution of POX 52. 
    Taking past theoretical predictions into account, the presence of the bulge would suggest that POX 52 underwent a galaxy merger(s), like massive galaxies. Also, considering that POX\,52 follows the $M_{\rm BH}$--$M_{\rm bulge}$ relation, observationally suggested by \cite{Sch19}, and also the possible past prolonged accretion, a moderate AGN feedback could have taken place. 

    \item Regarding the AGN structure, 
    the warm coronal properties ($\tau_{\rm warm}$ and $kT_{\rm e, warm}$) 
    are similar to those found for type-1 AGNs with larger BH masses. 
    The primary power-law emission believed to be originating in the hot corona is soft with $\Gamma \approx 2$, consistent with what is expected from the relation with the Eddington ratio. 
    Also, we have found that the relative strength to the bolometric luminosity ($2.2\times10^{43}$ \ergs), or the 2--10\,keV bolometric correction factor of $\approx$ 45, is consistent with an expected value from the 
    Eddington-ratio dependent relation \citep[e.g.,][]{Vas07,Vas09}, rather than that from a luminosity-dependent one \citep[e.g., ][]{Mar04,Dur20}. 
    Finally, we have found that the geometrically thin torus 
    constrained by the SED and X-ray data is consistent with the
    radiation pressure model, where the Eddington ratio is the key parameter. In summary, the corona-disk system is similar to those suggested for typical type-1 AGNs, and its nature as well as that of the torus seem to favor the Eddington-ratio-dependent scenarios. However, 
    to explain a considerable MIR deficit found in discussing the torus emission, an additional factor may be needed. Future observations of gas and dust at wavelengths longer than far-IR will be fundamental to fully understand the observed MIR deficit.  

    \item We have summarized the spectral data of the model AGN SED (Figure~\ref{fig:agnsed}) in Table~\ref{tab:agnsed}, including the data for each component. This information would be useful for discussing whether POX 52-like populations, therefore including rapidly accreting seed-BHs at high redshifts, can be detected with existing and future telescopes (e.g., {\it JWST}, {\it NewAthena}, {\it AXIS}, and {\it LUVOIR}).
    
\end{itemize}

We thank the reviewer for the useful comments, which helped us improve the quality of the manuscript.
T.K. and S.Y. are grateful for support from RIKEN Special Postdoctoral Researcher Program. 
T.K., S.Y., and H.N. are supported by JSPS KAKENHI grant numbers 
JP23K13153, 22K20391/23K13154, and 19K21884/20KK0071/20H0941/20H01947, respectively. 
C.R. acknowledges support from the Fondecyt Regular grant 1230345 and ANID BASAL project FB210003.
M.J.T. acknowledges support from FONDECYT Postdoctoral fellowship 3220516. 

This research has made use of the \nustar Data Analysis Software (NuSTARDAS) jointly developed by the ASI Space Science Data Center (SSDC, Italy) and the California Institute of Technology (Caltech, USA).

This work is based on observations obtained with the ESA science mission {\it XMM-Newton}, with instruments and contributions directly funded by ESA Member States and the USA (NASA), the {\it NuSTAR} mission, a project led by the California Institute of Technology, managed by the Jet Propulsion Laboratory and funded by NASA. 

This paper employs data, obtained by the Chandra X-ray Observatory, contained in~\dataset[Chandra Data Collection (CDC) 168]{
https://doi.org/10.25574/cdc.168}.

This work is based on observations made with the NASA Galaxy Evolution Explorer. GALEX is operated for NASA by the California Institute of Technology under NASA contract NAS5-98034\dataset[(10.17909/T9H59D)]{\doi{10.17909/T9H59D}}.

The Pan-STARRS1 Surveys (PS1) and the PS1 public science archive\dataset[(doi:10.17909/s0zg-jx37)]{\doi{doi:10.17909/s0zg-jx37}} have been made possible through contributions by the Institute for Astronomy, the University of Hawaii, the Pan-STARRS Project Office, the Max-Planck Society and its participating institutes, the Max Planck Institute for Astronomy, Heidelberg and the Max Planck Institute for Extraterrestrial Physics, Garching, The Johns Hopkins University, Durham University, the University of Edinburgh, the Queen's University Belfast, the Harvard-Smithsonian Center for Astrophysics, the Las Cumbres Observatory Global Telescope Network Incorporated, the National Central University of Taiwan, the Space Telescope Science Institute, the National Aeronautics and Space Administration under Grant No. NNX08AR22G issued through the Planetary Science Division of the NASA Science Mission Directorate, the National Science Foundation Grant No. AST-1238877, the University of Maryland, Eotvos Lorand University (ELTE), the Los Alamos National Laboratory, and the Gordon and Betty Moore Foundation.

This research is based on observations made with the NASA/ESA Hubble Space Telescope obtained from the Space Telescope Science Institute, which is operated by the Association of Universities for Research in Astronomy, Inc., under NASA contract NAS 5–26555. These observations are associated with program 10239\dataset[(doi:10.17909/j2ew-nh46)]{\doi{doi:10.17909/j2ew-nh46}}.

This publication makes use of data products from the 2MASS, which is a joint project of the University of Massachusetts and the Infrared Processing and Analysis Center/California Institute of Technology, funded by the National Aeronautics and Space Administration and the National Science Foundation.

This publication makes use of data products from the Wide-field Infrared Survey Explorer, which is a joint project of the University of California, Los Angeles, and the Jet Propulsion Laboratory/California Institute of Technology, funded by the National Aeronautics and Space Administration.
This publication also makes use of data products from NEOWISE, which is a project of the Jet Propulsion Laboratory/California Institute of Technology, funded by the Planetary Science Division of the National Aeronautics and Space Administration. 

This work is based in part on observations made with the Spitzer Space Telescope, which is operated by the Jet Propulsion Laboratory, California Institute of Technology under a contract with NASA. The Combined Atlas of Sources with Spitzer IRS Spectra (CASSIS) is a product of the IRS instrument team, supported by NASA and JPL. CASSIS is supported by the "Programme National de Physique Stellaire" (PNPS) of CNRS/INSU co-funded by CEA and CNES and through the "Programme National Physique et Chimie du Milieu Interstellaire" (PCMI) of CNRS/INSU with INC/INP co-funded by CEA and CNES.

We acknowledge the use of the NASA/IPAC Extragalactic Database (NED), which is operated by the Jet Propulsion Laboratory, California Institute of Technology, under contract with the National Aeronautics and Space Administration. 

This work has also made use of the VizieR catalogue access tool, CDS, Strasbourg, France.

\facilities{
\textit{XMM-Newton}, \textit{NuSTAR},  \textit{GALEX}, PanSTARRS, \textit{WISE}, \textit{Spitzer}, VLA, \textit{HST} 
}

\software{
CASA \citep{McM07}, 
NumPy \citep{Har20}, 
SciPy \citep{Vir20},
Pandas \citep{Tea22}, 
Matplotlib \citep{Hun07}, 
Astropy \citep{Ast13,Ast18}, 
CIGALE \citep{Boq19}, 
GALFIT \citep{Pen02}, 
HEAsoft (v6.30; Nasa High Energy Astrophysics Science Archive Research Center (Heasarc), 2014), 
NuSTARDAS (v2.0.0), 
SAS \citep[v18.0.0][]{Gab04}, 
CIAO \citep[v4.12;][]{Fru06},
XSPEC \citep[v12.12.1;][]{Arn96},
XCLUMPY \citep{Tan19}, 
RELXILL \citep[v1.4.0;][]{Dau13,Gar14} 
}


\bibliography{ref}
\bibliographystyle{aasjournal}

\end{document}